\documentclass[12pt]{article}

\def\fnote#1#2{\begingroup\def\thefootnote{#1}\footnote{#2}\addtocounter{footnote}{-1}\endgroup}

\usepackage{amssymb}

\def\inbar{\vrule height1.5ex width.4pt depth0pt}
\def\IB{\relax{\rm I\kern-.18em B}}
\def\IC{\relax\,\hbox{$\inbar\kern-.3em{\rm C}$}}
\def\ID{\relax{\rm I\kern-.18em D}}
\def\IE{\relax{\rm I\kern-.18em E}}
\def\IF{\relax{\rm I\kern-.18em F}}
\def\IG{\relax\,\hbox{$\inbar\kern-.3em{\rm G}$}}
\def\IH{\relax{\rm I\kern-.18em H}}
\def\II{\relax{\rm I\kern-.18em I}}
\def\IK{\relax{\rm I\kern-.18em K}}
\def\IL{\relax{\rm I\kern-.18em L}}
\def\IM{\relax{\rm I\kern-.18em M}}
\def\IN{\relax{\rm I\kern-.18em N}}
\def\IO{\relax\,\hbox{$\inbar\kern-.3em{\rm O}$}}
\def\IP{\relax{\rm I\kern-.18em P}}
\def\IQ{\relax\,\hbox{$\inbar\kern-.3em{\rm Q}$}}
\def\IR{\relax{\rm I\kern-.18em R}}
\def\IT{\relax{\rm I\kern-.18em T}}
\def\ZZ{\relax{\sf Z\kern-.4em Z}}

\def\a{\alpha}   \def\b{\beta}    \def\g{\gamma}  \def\d{\delta}
\def\e{\epsilon} \def\G{\Gamma}     \def\l{\lambda}
    \def\Om{\Omega} \def\si{\sigma}

\def\cA{{\cal A}} 
   
 \def\cH{{\cal H}}  
   
\def\cO{{\cal O}} \def\cP{{\cal P}}

\def\afrak{{\mathfrak a}} 
\def\mfrak{{\mathfrak m}} \def\pfrak{{\mathfrak p}}

\def\mathC{{\mathbb C}}
\def\mathF{{\mathbb F}}  \def\mathN{{\mathbb N}}
\def\mathP{{\mathbb P}}  \def\mathQ{{\mathbb Q}} \def\mathZ{{\mathbb Z}}

\def\balpha{{\bar \alpha}}
\def\bchi{{\bar \chi}}

 \def\Gwhat{{\widehat G}}

\def\fnote#1#2{\begingroup\def\thefootnote{#1}\footnote{#2}\addtocounter
{footnote}{-1}\endgroup}
\def\beq{\begin{equation}}
\def\eeq{\end{equation}}
\def\bea{\begin{eqnarray}}
\def\eea{\end{eqnarray}}
\def\llea#1{\label{#1}\eea}
\def\lleq#1{\label{#1}\eeq}
\let\nn=\nonumber
\def\tabroom{\hbox to0pt{\phantom{\Huge A}\hss}}
\def\notin{\ \hbox{{$\in$}\kern-.51em\hbox{/}}}

\def\lra{\longrightarrow}

\def\vphi{\varphi}

\def\un{{\underline n}}

\def\oa{{\overline a}}

\def\oC{{\overline C}}

\def\rmA{{\rm A}} \def\rmB{{\rm B}}
  
\def\rmI{{\rm I}} \def\rmK{{\rm K}}
 \def\rmN{{\rm N}} 

     \def\rmdeg{{\rm deg}}
   \def\rmdim{{\rm dim}}

   \def\rmmod{{\rm mod}}
 
\def\rmpt{{\rm pt}}      \def\rmres{{\rm res}}
\def\rmsign{{\rm sign}}
\def\rmth{{\rm th}}

\def\rmCorr{{\rm Corr}}
        \def\rmGal{{\rm Gal}}

\def\rmHom{{\rm Hom}} \def\rmHW{{\rm HW}}
\def\rmII{{\rm II}}   \def\rmIII{{\rm III}}     \def\rmIV{{\rm IV}}

\def\rmSL{{\rm SL}}      
\def\rmSym{{\rm Sym}}

\def\notdiv{{\relax{~|\kern-.35em /~}}}

\def\XFq{{X/\IF_q}}

\def\boxit#1{
\vbox{\hrule height1pt\hbox{\vrule width1pt\kern0.3cm
\vbox{\kern0.3cm\hbox{$\displaystyle#1$}\kern0.3cm}\kern0.3cm\vrule
width1pt}\hrule height1pt}}

\begin{document}
\parindent=0pt

\phantom{\bf Draft}


\vskip 1truein

 \centerline{\Large {\bf Emergent Spacetime from Modular Motives}}

\vskip .3truein

\centerline{\sc Rolf Schimmrigk \fnote{$\dagger$}{email:
netahu@yahoo.com}}

\vskip .3truein

 \centerline{\it Indiana University South Bend}
 \vskip .05truein
 \centerline{\it 1700 Mishawaka Ave., South Bend, IN 46634}

\vskip 1truein

\baselineskip=17pt

\centerline{\bf ABSTRACT:}

\vskip .1truein

\begin{quote}

 The program of constructing spacetime geometry from string theoretic
 modular forms is extended to Calabi-Yau varieties of dimensions two, three,
 and four, as well as higher rank motives. Modular forms on the worldsheet
 can be constructed from the geometry of spacetime by computing the
 L-functions associated to omega motives of Calabi-Yau varieties,
 generated by their holomorphic $n-$forms via Galois representations.
 The modular forms that emerge from the $\Om-$motive and other motives of
 the intermediate cohomology are related to characters
 of the underlying rational conformal field theory. The converse
 problem of constructing space from string theory proceeds in the class
 of diagonal theories by determining the motives associated to modular
 forms in the category of pure motives with complex multiplication.
 The emerging picture indicates that the L-function can be
 interpreted as a map from the geometric category of motives
 to the category of conformal field theories on the worldsheet.
\end{quote}

\vfill

{\sc PACS Numbers and Keywords:} \hfill \break Math:  11G25
Varieties over finite fields; 11G40 L-functions; 14G10  Zeta
functions; 14G40 Arithmetic Varieties \hfill \break Phys: 11.25.-w
Fundamental strings; 11.25.Hf Conformal Field Theory; 11.25.Mj
Compactification

\renewcommand\thepage{}
\newpage
\parindent=0pt

\baselineskip=17pt

\pagenumbering{arabic}

\tableofcontents


 \vfill \eject

\baselineskip=18pt

\parskip=.13truein

\section{Introduction}

The present paper continues the program of applying methods from
arithmetic geometry to the problem of understanding the question
how spacetime emerges in string theory.  The goal is to construct
a direct relation between the physics on the worldsheet and the
geometry of the extra dimensions. One way to formulate this
question is by asking whether it is possible to explicitly
determine the geometry of the compact dimensions from the building
blocks of the two-dimensional string structure. In this general,
but vague, form the problem of constructing an emergent geometry
in string theory could have been formulated more than thirty years
ago. The reason that it was not can probably be traced to both the
lack of a concrete framework, and the lack of useful tools. The
framework of the heterotic string of the 1980s, in combination
with the web of dualities between different string models
discovered the 1990s, motivates a more concrete version of this
problem, which aims at the relation between Calabi-Yau varieties
and worldsheet physics given by exactly solvable conformal field
theories. Both, Calabi-Yau varieties and exactly solvable field
theories define rich structures, raising a number of problems
which have not been addressed in the past.

The key ingredient of the program pursued here is the modular
invariance of the theory. From a spacetime physics perspective it
is initially somewhat surprising that this feature of string
theory should turn out to provide a useful tool for the
understanding of its geometric consequences, because it is the
modular invariance of the two-dimensional theory that a priori
appears most difficult to explain from a geometric perspective. By
now there exists a fair amount of evidence that shows that methods
from arithmetic geometry provide promising tools for this problem,
at least in lower dimensions. The main purpose of the present
paper is to generalize previous results by constructing a class of
motives for all Calabi-Yau manifolds (and Fano varieties of
special type), independent of any specific construction, and to
analyze their modularity properties in the context of weighted
Fermat varieties (manifolds of Brieskorn-Pham type). As a
consequence, string theoretic modularity emerges for varieties of
dimensions three and four, relevant for string, M- and F-theory,
including motives of higher rank).

 The basic problem of extending modularity results for L-functions
 in dimensions larger than one is made difficult by the fact
 that no generalization of the
 elliptic modularity theorem \cite{w95, bcdt01} is known,
 even conjecturally. This makes even the first step, of
 constructing modular forms from algebraic varieties, nontrivial.
There exists, however, a program, associated most closely with the
name Langlands, that suggests that even in higher dimensions the
Hasse-Weil L-functions of geometric structures have modular
properties in a generalized sense. It is expected in particular
that associated to each cohomology group is an automorphic
representation, leading to an automorphic form.
  The class of automorphic L-functions contains a special type of objects,
  the standard L-functions, which
   generalize the Hecke L-functions \cite{rl78}.
Modularity of Hecke L-functions is known in virtue of their
analytic continuation and their functional equations
\cite{h36,w67}. Langlands' vision thus is based on results
obtained by Artin and Hecke. While Artin considered
representations $\rho$ of the Galois group $\rmGal(K/\mathQ)$ of a
number field $K$ to define L-functions $L(\rho,s)$, Hecke had
previously introduced L-functions based on certain characters
$\chi$ associated to number fields (called Gr\"o\ss encharaktere
by Hecke, also called algebraic Hecke characters), whose structure
was motivated by an attempt to establish modularity of the
L-function. It turned out that these a priori different concepts
lead to the same object in the sense that Artin's L-functions and
Hecke's L-functions agree \cite{rl70, rl71}. Langlands' conjecture
involves a generalization of the Artin-Hecke framework to GL$(n)$.
More precisely, the connection between geometry and arithmetic can
be made because representations of the Galois group can be
constructed by considering the $\ell-$adic cohomology as a
representation space. This strategy has proven difficult to
implement for higher $n$ in general, and in particular in the
context of obtaining a string theoretic interpretation of
geometric modular forms beyond the case of elliptic curves and
rigid Calabi-Yau varieties.

For varieties of higher dimensions the results obtained so far
indicate that it is more important to identify irreducible pieces
of low rank in the cohomology groups, and to consider the
L-functions of these subspaces. The difficulty here is that at
present there exists no general framework that provides guidance
for the necessary decomposition of the full cohomology groups.
Nevertheless, the Langlands program suggests that modularity, and
more generally automorphy, are phenomena that transcend the
framework of elliptic curves, and one can ask the question whether
the methods described in \cite{su02, ls04, rs05} to establish
modularity relations between elliptic curves and conformal field
theories can be generalized to higher dimensional varieties.
Results in this direction have been obtained for extremal K3
surfaces of Brieskorn-Pham type in ref. \cite{rs06}.

In the present paper the string modularity results obtained
previously are extended to all higher dimensions that are of
physical relevance. The idea is to consider particular subgroups
of the intermediate cohomology group of a variety, defined by the
representation of the Galois group associated to the manifold. For
manifolds of Calabi-Yau and special Fano type there exists at
least one nontrivial orbit, defined by the holomorphic $n-$form in
the case of Calabi-Yau spaces, and the corresponding cohomology
group in the case of special Fano manifold. This orbit will be
called the omega motive. The strategy developed here is completely
general and can be applied to any Calabi-Yau variety, as well as
Fano varieties of special type. In later sections the framework
developed will be applied to Calabi-Yau varieties of
Brieskorn-Pham type, the class of varieties for which Gepner
\cite{g87} originally discovered a relation between the spectra of
a certain type of conformal field theory and the cohomology of the
manifolds. For this class the $\Om-$motive, as well as the other
submotives, can be determined explicitly.

A simplifying characteristic of the class of extremal K3 surfaces
of Brieskorn-Pham type considered in \cite{rs06} is that their
$\Om-$motive is of rank two. This is not the case in general, and
it is therefore of interest to see whether it is possible to
extend the string analysis of \cite{rs06} to K3 motives of higher
rank.

Define the K3 surface
 \beq
  X_2^{12} =
  \left\{(z_0:z_1:z_2:z_3) \in \mathP_{(2,3,3,4)}~{\Big |}~
   z_0^6 + z_1^4 + z_2^4 + z_3^3=0\right\}
 \lleq{k3deg12}
 and denote by $E^4 \subset \mathP_{(1,1,2)}$ and $E^6 \subset
 \mathP_{(1,2,3)}$ the two weighted Fermat curves of degree four
 and six, respectively \cite{rs05}.
The following result is shown.

{\bf Theorem 1.}~ {\it The L-series of $X_2^{12}$ is given by the
Mellin transform of product of the modular forms $f_{E^4}$ and
$f_{E^6}$ of the elliptic curves $E^4$ and $E^6$}
   \beq
   L_{\Om}(X_2^{12},s) = L(f_{E^4}\otimes f_{E^6},s).
  \eeq
The precise meaning of the tensor product of modular forms will
become clear below.

In dimension three consider the manifolds
 \bea
  X_3^6 &=& \{z_0^6 + z_1^6 + z_2^6 + z_3^6 + z_4^3 =0\} \subset
           \mathP_{(1,1,1,1,2)} \nn \\
  X_3^{12} &=& \{z_0^6 + z_1^6 + z_2^6 + z_3^4 + z_4^4 = 0\} \subset
           \mathP_{(2,2,2,3,3)}
  \llea{modular-3folds}
and denote by $f_{X_2^{\rm 6A}}$ the modular form of the
$\Om-$motive of the weighted Fermat K3 surface $X_2^{\rm 6A}
\subset \mathP_{(1,1,1,3)}$, determined in \cite{rs06} and
described in Section 6. The following results hold.

{\bf Theorem 2.} \hfill \break
 {\it 1) The inverse Mellin transform of the $\Om-$motivic $L_{\Om}(X_3^6,s)$ of
 the threefold $X_3^6$ is a cusp form $f_\Om(q)$ of weight four and level 108
 for Hecke's congruence subgroup. The
 L-function of the intermediate cohomology group $H^3(X_3^6)$ decomposes
 into modular factors as}
 \beq
 L(H^3(X_3^6),s) = L(f_\Om,s) \cdot \prod_i L(f_i\otimes
 \chi_i,s),
 \eeq
 {\it where $f_i\in S_2(\G_0(N_i))$ with $N_i=27,144,432$, and
 $\chi_i$ are twist characters (that can be trivial).
 \hfill \break
 2) The $\Om-$motivic L-series of $X_3^{12}$ is the Mellin transform
 of the modular forms associated to $E^4 \subset \mathP_{(1,1,2)}$ and
 the extremal weighted Fermat surface $X_2^{6\rmA} \subset \mathP_{(1,1,1,3)}$}
  \beq
   L_{\Om}(X_3^{12},s) = L(f_{E^4}\otimes f_{X_2^{6\rmA}},s).
  \eeq
  {\it The L-series of the remaining part of the intermediate
  cohomology $H^3(X_3^{12})$ decomposes into a product of factors
  that include L-series of modular forms of weight two and levels
  $N=27,64,144, 432$, possibly including a twist.}

As a last example the modularity of the $\Om-$motive of the degree
six fourfold
  \beq
  X_4^6 = \left\{(z_0:\cdots :z_5)\in \mathP_5~{\Big |}~
     \sum_{i=0}^5 z_i^6 = 0 \right\}
 \lleq{deg6fourfold}
 is shown to be modular.

{\bf Theorem 3.}~ {\it The inverse Mellin transform of the
  L-function of the $\Om-$motive of $X_4^6$ is of the form}
  \beq
  f_{\Om}(X_4^6,q) = f_{27}(q)\otimes \chi_3
  \eeq
  {\it where $f_{27}(q)$ is a cusp Hecke
  eigenform of weight $w=5$ and level $N=27$,
  and $\chi_3$ is the Legendre character. There exists
  an algebraic Hecke character $\psi_{27}$ with congruence ideal
  $\mfrak = (3)$ such that the motivic L-series is its
  $L_{\Om}(X_4^6,s) = L(\psi_{27}^4,s)\otimes
  \chi_3$.}

The basic question raised by these results is whether the
$\Om-$motive is string automorphic in general. In a larger
context, one may view the L-function as a link between the
geometry of spacetime and the physics of the worldsheet. One way
to make this idea more explicit is by viewing $L$ as a functor
from the category of Fano varieties of special type (or rather
their motives) to the category of superconformal field theories.
The evidence obtained so far supports this perspective for a
physical interpretation of L-functions. The notion e.g. of
composing motives then translates into a corresponding composition
of conformal field theories. A concrete example is the motivic
tensor structure which maps into a tensor structure for conformal
field theories. The basic tensor structure of motives is described
in L-function terms by the Rankin-Selberg convolution
$L(f_1\otimes f_2,s)$ of the modular forms $f_i$ of the modular
motives $M_i$, and also leads to the symmetric square of modular
forms. Denote by $\Theta_\chi = \prod_i \Theta_i \otimes \chi$
twisted products constructed from modular forms $\Theta_i$ on the
string worldsheet. The motivic L-functions $L(M(X),s)$ that emerge
from Calabi-Yau varieties and special Fano varieties can be
expressed in terms of string modular L-functions of the type
 \beq
  L(\Theta_\chi,s),~~~
  L(\rmSym^r \Theta_\chi,s),~~~
  L(\Theta_{\chi_1}^1 \otimes \Theta_{\chi_2}^2,s),
 \eeq
 where $L(\rmSym^r f,s)$ describes the L-function associated to a
 symmetric tensor product of a modular form $f$.

The paper is organized as follows. Section 2 briefly introduces
the necessary modular theoretic background. Section 3 describes
the notion of a Grothendieck motive and defines the concept of
 $\Om-$motives for an arbitrary Calabi-Yau manifold, as well as
 for the class of Fano varieties of special type. This provides
 the framework for the relation between the geometry of spacetime and
 physics on the string worldsheet. Sections 4 and 5 describe the basic
 structure of the L-functions for Calabi-Yau surfaces and threefolds,
 derived from Artin's zeta function. Sections 6 briefly reviews the results for
modular motives of rank two that appear as building blocks for the
examples of higher dimension and higher rank described in Sections
7 through 11.  Section 12 shows how the converse problem can be
approached in the context of modular motives. Section 13 ends the
paper with some final remarks.

\vskip .2truein

\section{Modularity}

In order to make the paper more self-contained the paragraphs
briefly summarize the types of modular forms that eventually are
reflected in the geometry of weighted hypersurfaces of Calabi-Yau
and special Fano type. The affine Lie algebraic forms introduced
by Kac and Peterson provide the structures on the worldsheet,
while certain types of modular Hecke L-series will arise from the
arithmetic of the geometry.

\subsection{Modular forms from affine Lie algebras}

The simplest class of N=2 supersymmetric exactly solvable theories
is built in terms of the affine SU(2) theory as a coset model
 \begin{equation}
 \frac{{\rm SU(2)}_k \otimes {\rm U(1)}_2}{{\rm U(1)}_{k+2,{\rm diag}}}.
 \end{equation}
 Coset theories $G/H$ lead to central charges of the form $c_G - c_H$,
 hence the
supersymmetric affine theory at level $k$  still has central
charge $c_k=3k/(k+2)$. The spectrum of anomalous dimensions
$\Delta^k_{\ell,q,s}$ and U(1)$-$charges $Q^k_{\ell,q,s}$ of the
primary fields $\Phi^k_{\ell,q,s}$ at level $k$ is given by
 \begin{eqnarray}
    \Delta^k_{\ell,q,s} &=& \frac{\ell (\ell +2)-q^2}{4(k+2)}
                             + \frac{s^2}{8} \nonumber \\
    Q^k_{\ell,q,s} &=& - \frac{q}{k+2} + \frac{s}{2},
 \end{eqnarray}
where $\ell\in \{0,1,\dots,k\}$, $\ell+q+s \in 2\mathbb{Z}$, and
$|q-s|\leq \ell$. Associated to the primary fields are characters
defined as
 \bea
 \chi^k_{\ell,q,s}(\tau, z,u)
  &=& e^{-2\pi i u} {\rm tr}_{\cH^{\ell}_{q,s}}
        e^{2\pi i\tau (L_0 -\frac{c}{24})} e^{2\pi i J_0} \nn \\
  &=& \sum c^k_{\ell,q+4j-s}(\tau) \theta_{2q+(4j-s)(k+2),2k(k+2)}(\tau, z,u),
 \eea
 where the trace is to taken over a projection $\cH^{\ell}_{q,s}$ to a
definite fermion number (mod 2) of a highest weight representation
of the (right-moving) $N=2$ algebra with highest weight vector
determined by the primary field. The expression of the rhs in
terms of the string functions \cite{kp84}
 \beq
 c^k_{\ell,m}(\tau) = \frac{\Theta^k_{\ell,m}(\tau)}{\eta^3(\tau)}
 \eeq
 where $\eta(\tau)$ is the Dedekind eta function, and
 $\Theta^k_{\ell,m}(\tau)$ are the Hecke indefinite modular
 forms
 \beq
 \Theta^k_{\ell,m}(\tau)
  = \sum_{\stackrel{\stackrel{-|x|<y\leq |x|}{(x,y)~{\rm
or}~(\frac{1}{2}-x,\frac{1}{2}+y)}}{\in
\ZZ^2+\left(\frac{\ell+1}{2(k+2)},\frac{m}{2k}\right)}} \rmsign(x)
e^{2\pi i \tau((k+2)x^2-ky^2)}
 \eeq
 and theta functions
 \beq
 \theta_{n,m}(\tau,z,u) = e^{-2\pi i m u} \sum_{\ell \in \ZZ +
 \frac{n}{2m}} e^{2\pi i m \ell^2 \tau + 2\pi i \ell z}.
 \eeq
 is useful because it follows from this
representation that the modular behavior of the $N=2$ characters
decomposes into a product of the affine SU(2) structure in the
$\ell$ index and into $\Theta$-function behavior in the charge and
sector index. It follows from the coset construction that the
essential ingredient in the conformal field theory is the SU(2)
affine theory.

The issue of understanding emergent spacetime in string theory can
now be reformulated in a more concrete way as as the problem of
relating string theoretic modular forms to geometric ones.  It
turns out that more important than the string functions are the
associated SU(2) theta functions $\Theta^k_{\ell,m}(\tau)$.
 These indefinite Hecke forms are associated to quadratic number fields
determined by the level of the affine theory. They are modular
forms of weight 1 and cannot, therefore, be identified with
geometric modular forms. It turns out, however, that appropriate
products lead to interesting motivic forms \cite{su02, ls04, rs05,
rs06}.

\subsection{Modular forms from algebraic Hecke characters}

The modularity of the L-series determined in this paper follows
from the fact that they can be interpreted in terms of Hecke
L-series associated to Gr\"o\ss encharaktere, defined by Jacobi
sums. A Gr\"o\ss encharakter, or algebraic Hecke character) can be
associated to any number field. Hecke's modularity discussion
\cite{h37b} has been extended by Shimura \cite{gs71b} and Ribet
\cite{r77}.

Let $K$ be a number field and
  $\sigma:~K ~\longrightarrow ~{\mathbb C}$ denote an embedding.

{\bf Definition.}~ {\it A Gr\"o\ss encharakter is a homomorphism
$\psi: I_{{\mathfrak m}}(K) \rightarrow {\mathbb C}^{\times}$ from
the fractional ideals of $K$ prime to the congruence integral
ideal ${\mathfrak m}$ such that
   $\psi((z)) = \sigma(z)^{w-1}$, for all $z\in K^{\times}$ such that
   $z\equiv 1({\rm mod}^{\times}{\mathfrak m})$. The type
   of behavior of $\psi$ on
   the principal ideals is called the infinity type.}

In the present case the cyclotomic Jacobi sums determined by the
finite field Jacobi sums computed above arise from imaginary
quadratic fields $K=\mathQ(\sqrt{-D})$, where where $-D$ is the
discriminant of the field. In this case the structure of these
characters simplifies. Denote by $\psi$ an algebraic Hecke
character of $K$ and by ${\rm N}{\mathfrak p}$ the norm of prime
ideal ${\mathfrak p}$ in the ring of integers $\cO_K$. The Hecke
L-series of $\psi$ is defined by
 \begin{equation}
  L(\psi,s) = \prod_{{\mathfrak p} \in {\rm Spec}~{\mathcal O}_K}
  \frac{1}{1-\frac{\psi({\mathfrak p})}{{\rm N}{\mathfrak p}^s}}.
  \end{equation}

 The modularity of the corresponding $q-$series
$f(\psi,q)=\sum_n a_nq^n$ associated to the L-series
  via the Mellin transform is characterized by a Nebentypus
  character $\e$ defined in terms of
  the Dirichlet character $\vphi$ associated to $K$ and
  a second Dirichlet character $\lambda$ defined mod
   ${\rm N}{\mathfrak m}$ by
  \begin{equation}
   \l(a) = \frac{\psi((a))}{\sigma(a)^{w-1}},\phantom{gap}a\in {\mathbb Z}.
  \end{equation}
  The Nebentypus character $\epsilon$ is given by the product
  $\epsilon = \chi \lambda$ of these two characters. Modularity
  of the L-series follows from the following result of Hecke.

  {\bf Theorem 4.}~ {\it Let $\psi$ be a Gr\"o\ss encharakter of the imaginary
  quadratic field $K$ with infinity type $\si^{w-1}$. Define the
   coefficients $c_n$ as}
   \beq
    \sum_{\stackrel{(\afrak,\mfrak)=1}{\afrak ~{\rm integral}}}
    \psi(\afrak)q^{\rmN\afrak} =: \sum_{n=1}^{\infty}c_nq^n.
   \eeq
   {\it Then there exists a unique newform $f=\sum_{n=1}^{\infty}
   a_nq^n$ of weight $w$ and character $\e = \l \vphi$
  such that}
     \beq
      a_p = c_p~~~~~~~\forall~p\notdiv D\rmN\mfrak.
     \eeq

Of particular importance in this paper are algebraic Hecke
characters associated to the Gauss field $\mathQ(\sqrt{-1})$ and
the Eisenstein field $\mathQ(\sqrt{-3})$. For $\mathQ(\sqrt{-1})$
consider prime ideals $\pfrak = (z_{\pfrak})$ and define the
character $\psi_{32}$ by setting
 \beq
  \psi_{32}(\pfrak) = z_{\pfrak}
 \eeq
 where the generator $z_{\pfrak}$ is determined by the congruence relation
 \beq
   z_{\pfrak} \equiv 1(\rmmod~(2+2i))
 \eeq
 for the congruence ideal $\mfrak = (2+2i)$.

 For the field $\mathQ(\sqrt{-3})$ two characters $\psi_N$, $N=27,
 36$ and their twists will appear. The congruence ideals here
 are given by
  \bea
    \mfrak_{27} &=& (3) \nn \\
    \mfrak_{36} &=& 1 + 2\xi_3
  \llea{eisenstein-hecke-chars}
 leading to the characters
  \beq
   \psi_N(\pfrak) = z_{\pfrak}
  \eeq
  where the generator is determined uniquely by the congruence
 relations
  \beq
    z_{\pfrak} \equiv 1(\rmmod_N).
  \eeq

\subsection{Rankin-Selberg products of modular forms}

An important ingredient in the analysis of higher dimensional
varieties is the fact that the L-series of motives carrying higher
dimensional representations of the Galois group can sometimes be
expressed in terms of L-functions of lower rank motives. This
leads to Rankin-Selberg products of L-functions. This construction
is quite general, but will be applied here only to products of
L-series that are associated to Hecke eigenforms. If the L-series
of two modular forms $f,g$ of arbitrary weight and arbitrary
levels are given by
 \bea
  L(f,s) &=& \sum_n a_n n^{-s} \nn \\
  L(g,s) &=& \sum_n b_n n^{-s},
 \eea
 it is natural to consider the naive Rankin-Selberg L-series
 associated to $f,g$ as
  $$
  L(f\times g,s) = \sum_n a_nb_nn^{-s}.
 $$
 It turns out that a slight modification of this product defined
 has better properties, and is more appropriate for geometric constructions.
 If $f \in S_{w_1}(\G_0(N),\e)$ and $g\in S_{w_2}(\G_0(N),\l)$ are cusp forms with
 characters $\e$ and $\l$, respectively, the modified Rankin-Selberg product is
 defined as
 \beq
  L(f\otimes g,s) = L_N(\e \l,2s+2-(w+v)) L(f\times g,s),
 \eeq
 where $L_N(\chi,s)$ is the truncated Dirichlet L-series defined
 by the condition that $\chi(n)=0$ if $(n,N)>1$ \cite{gs76}.

 Hecke showed that such forms $f,g$ have Euler products given by
 \bea
  L(f,s) &=& \prod_p [(1-\a_p p^{-s})(1-\b_pp^{-s})]^{-1} \nn \\
  L(g,s) &=& \prod_p [(1-\g_p p^{-s})(1-\d_pp^{-s})]^{-1},
 \eea
 where $\a_p+\b_p = a_p$, $\g_p+\d_p=b_p$ and $\a_p\b_p =
 p^{w-1}$, $\g_p\d_p=p^{v-1}$.
 It can be shown that the modified Rankin-Selberg product has the Euler product
 \beq
 L(f\otimes g,s) = \prod_p
 [(1-\a_p\g_p p^{-s})(1-\a_p\d_p p^{-s})
   (1-\b_p\g_pp^{-s})(1-\b_p\d_p p^{-s})]^{-1}.
 \eeq
 The tensor notation is at this point formal, but it will become
 clear that this product is indicative of a representation
 theoretic tensor product.  Furthermore, it also describes the L-series
 of the tensor product $M_f\otimes M_g$ of motives $M_f, M_g$
 associated to the modular forms $f,g$ via the constructions of
 Deligne \cite{d69}, Jannsen \cite{j90} and Scholl \cite{as90}
 $$
 L(f\otimes g,s) ~=~ L(M_f\otimes M_g,s).
 $$
 The motivic tensor product will be described below.

 The Rankin-Selberg products which will appear below involve
 modular forms of weight two and three, leading to rank four
 motives on Calabi-Yau varieties of dimension two and three.

\subsection{Complex multiplication modular forms}

A special class of modular forms that is relevant in this paper
are forms which are sparse in the sense that a particular subset
of the coefficients $a_p$ of their Fourier expansion $f(q) =
\sum_n a_nq^n$ vanish. A conceptual way to formulate was
introduced by Ribet \cite{r77}. A complex multiplication (CM)
modular form $f(q)$ is characterized by the existence of an
imaginary quadratic number field $K=\mathQ(\sqrt{-d})$ such that
the coefficients $a_p$ vanish for those rational primes $p$ which
are inert in $K$. If follows from this definition that any such
form can be described by the inverse Mellin transform of an
L-series associated to a Hecke Gr\"o\ss encharakter, which is the
view originally adopted by Hecke, and also Shimura.

Modular forms with complex multiplication are more transparent
than general forms, in particular in the context of their
associated geometry. This will become important further below in
the construction of the Calabi-Yau motives from the conformal
field theory on the worldsheet.

\vskip .2truein

\section{$\Om-$Motives}

The results in the present and previous papers show that it is
useful for modularity to consider the Galois orbit in the
cohomology defined by the holomorphic $n-$form in a Calabi-Yau
variety. This Galois orbit defines a geometric substructure of the
manifold, called the $\Om-$motive, which is an example of a
Grothendieck motive on these manifolds. Similar orbits can be
considered in the context of so-called special Fano varieties
considered in \cite{rs92, cdp93, rs94}, whose modularity
properties are analyzed in the context of mirror pairs of rigid
Calabi-Yau varieties in \cite{kls08}. The aim of the present
section is to provide the background for the reconstruction of the
motivic structure from the conformal field theory. The circle of
ideas that is concerned with the relations between characters,
modular forms, and motives extends beyond the class of weighted
Fermat varieties, and it is useful to formulate the constructions
in a general framework.

The outline of this section is to first describe the concept of
Grothendieck motives, also called pure motives, then to define the
notion of $\Om-$motives in complete generality, and finally to
consider the $\Om-$motive in detail for weighted Fermat
hypersurfaces, i.e. of Brieskorn-Pham type.

\subsection{Grothendieck motives}

The idea to construct varieties directly from the conformal field
theory on the worldsheet without the intermediary of
Landau-Ginzburg theories or sigma models becomes more complex as
the number of dimensions increases. Mirror symmetry and other
dualities show that it should not be expected that any particular
model on the worldsheet should lead to a unique variety. Rather,
one should view manifolds as objects which can be build from
irreducible geometric structures. This physical expectation is
compatible with Grothendieck's notion of motives. The original
idea for the existence of motives arose from a plethora of
cohomology theories Grothendieck was led to during his pursuit of
the Weil conjectures \cite{m68,md69,k72}. There are several ways
to think of motives as structures that support these various
cohomology theories, such as Betti, de Rham, \'etale, crystalline
cohomology groups etc., and to view these cohomology groups as
realization of motives.

Grothendieck's vision of motives as basic building blocks that
support universal structures is based on the notion of
correspondences. This is an old concept that goes back to Klein
and Hurwitz in the late 19th century. The idea is to define a
relation between two varieties by considering an algebraic cycle
class on their product. In order to do so an algebraic cycle is
defined as a finite linear combination of irreducible subvarieties
$V_{\a} \subset X$ of codimension $r$ of a variety $X$. The set of
all these algebraic cycles defines a group
 \beq
 Z^r = \left\{\sum_{\a} n_{\a}V_{\a}~{\Big |}~V_{\a} \subset X.
  \right\}
 \eeq
 This group is too large to be useful, hence one considers
 equivalence relations between its elements. There are a variety
 of such equivalence relations, resulting in quite different
 structures. The most common of these are rational, homological,
 and numerical equivalence. A description of these can be found in
 \cite{m96}, but for the following it will not be
 important which of these is chosen. Given any of these
 equivalences one considers the group of equivalence classes
 \beq
  A^r(X) = Z^r(X)/\sim
 \eeq
  of
 algebraic cycles to define the group of
  algebraic correspondences of degree $r$
 between manifolds $X,Y$ of equal dimension $d$ as
 \beq
 \rmCorr^r(X,Y) = A^{d+r}(X\times Y).
 \eeq
 Correspondences can be composed $f\cdot g$, leading to the notion of a
 projector $p$ such that $p\cdot p=p$. The first step in the
 construction of Grothendieck motives is the definition of an
 effective motive, obtained by considering a pair defined by a
 variety and a projector $M=(X,p)$, where $p$ is a projector in
 the ring of algebraic correspondences of degree zero, $p\in
 \rmCorr^0(X,X)$. Maps between such objects are of the form
  \cite{j92}
  \beq
  \rmHom((X,p),(Y,q)) ~= ~
   q\circ \rmCorr^0(X,Y)_{\mathQ}\circ p.
  \eeq
 This formulation of morphisms between effective motives is
 equivalent to the original view of Grothendieck described in
 \cite{k72}.

 It is important for physical applications of motives
  to enlarge the class of effective
 motives by introducing twists of effective motives by powers of
 the inverse Lefschetz motive $L$. This is an effective motive
  defined as $L=(\mathP_1,1-Z)$, where $Z$ is
 the cycle class $Z\in A^1(\mathP_1\times \mathP_1)$ defined by
 the cycle $\mathP_1\times \rmpt$. It is possible to tensor
 effective motives $M$ by $L$ and its inverse. Combining the
 notions of effective motives and the Lefschetz motive leads to
 the concept of a Grothendieck motive.

 {\bf Definition.}~{\it A Grothendieck motive is a triple
 $M(m)=(X,p,m)$, where $M=(X,p)$ is an effective motive, and
 $m\in \mathZ$. $M(m)=(M,m)$ is the $m-$fold Tate twist of $M$.
 If $N(n)=(Y,q,n)$ is another motive morphisms are defined as}
  \beq
  \rmHom(M(m),N(n)):= p\circ \rmCorr^{n-m}(X,Y)\circ q.
  \eeq

The tensor product of two Grothendieck motives $M_i=(X_i,p_i,m_i),
i=1,2$ is defined as
 \beq
  M_1 \otimes M_2 = (X_1\times X_2, p_1\oplus p_2, m_1+m_2).
 \eeq

 A discussion of the virtues and
disadvantages of the various realizations in terms of specific
equivalence relations can be found in \cite{j92, as94}, building
on earlier references, such as \cite{m68,md69,k72}. A more
detailed discussion of motives can be found in \cite{jks94}.

\subsection{$\Om-$motives for manifolds of Calabi-Yau and special Fano type}

When considering the emergent geometry problem via string
theoretic modular forms it is of interest to consider
 L$-$functions associated to motives of low rank, not of
  the full cohomology groups of a variety. For higher genus curves and higher
dimensional varieties the experimental evidence \cite{su02, ls04,
rs05, rs06} suggests that the relevant physical information is
encoded in subspaces of the cohomology.  A possible strategy
therefore is to consider the factorization of L$-$functions and to
ask whether modular forms arise from the emerging pieces, and if
so, whether these modular forms admit a string theoretic Kac-Moody
 interpretation.

This section describes a general strategy, valid for any
Calabi-Yau variety $X$ of dimension $\rmdim_{\mathC}X=d$, for
decomposing the L-function of its intermediate cohomology
$H^d(X)$, into pieces which lead to L-functions with integral
coefficients. These L-functions then have the potential, when
modular, to admit a factorization into Kac-Moody theoretic modular
forms along the lines discussed in \cite{su02, ls04, rs05} for
elliptic curves and higher genus curves. The basic strategy
outlined below is a generalization of the method described in
\cite{rs06}, which was based on Jacobi sums associated to
hypersurfaces embedded in weighted projective spaces.

The idea is to consider an orbit in the cohomology which is
generated by the holomorphic $d-$form $\Om \in H^{d,0}(X)$ via the
 action of the Galois group $\rmGal(K/\mathQ)$ of the number field
 determined by the arithmetic properties of the variety as
 dictated by the Weil conjectures \cite{w49}
 proven by Grothendieck \cite{g65} and
 Deligne \cite{d74}. The resulting orbit of this action turns out
 to define a motive in the sense of Grothendieck, as will be shown
 further below. To see how the group structure appears in full
 generality it is necessary to briefly review the arithmetic
 structure of arbitrary Calabi-Yau varieties.

For a general smooth algebraic variety $X$ reduced mod $p$ the
congruence zeta function of $\XFq$ is defined by
 \beq
  Z(X/\mathF_q, t) \equiv \exp\left(\sum_{r\in \mathN} \#
      \left(X/\mathF_{q^r} \right) \frac{t^r}{r}\right).
  \eeq
  Here the sum is over all finite extensions
  $\mathF_{q^r}$ of $\mathF_q$ of degree
  $r = \left[\mathF_{q^r}:\mathF_q\right]$.
 Per definition $Z(X/\mathF_q,t) \in 1 + \mathQ[[t]]$, but the
 expansion can be shown to be integer valued by writing it as an
 Euler product. The main virtue of $Z(X/\mathF_q, t)$ is that the
 numbers $N_{p^r}= \# \left(X/\mathF_{p^r}\right)$ show a simple
 behavior, as a result of which
the zeta function can be shown to be a rational function.

{\bf 1)} The first step is to consider the rational form of
Artin's congruent zeta function. This leads to a link between the
purely arithmetic geometric objects $N_{r,p}=\#(X/\mathF_{p^r})$
for all primes $p$ and the cohomology of the variety. This was
first shown for curves by F.K. Schmidt in the thirties in letters
to Hasse \cite{fks31}\cite{hh33}\cite{hh34}. Further experience by
Hasse, Weil, and others led to the conjecture that this phenomenon
is more general, culminating in the cohomological part of Weil
conjecture. According to Weil \cite{w49} and Grothendieck
\cite{g65}  $ Z(X/\mathF_p,t)$ is a rational function which can be
written as
 \beq
Z(X/\mathF_p,t)
 =\frac{\prod_{j=1}^d
 \cP_p^{2j-1}(t)}{\prod_{j=0}^d \cP_p^{2j}(t)},
 \lleq{groth65}
  where
  \beq
   \cP^0_p(t)=1-t,~~~ \cP^{2d}_p(t)=1-p^dt
 \eeq
  and for $1\leq j \leq 2d-1$
  \beq
 \rmdeg~\cP_p^j(t) = b^j(X),
  \eeq
  where $b^j(X)$ denotes the $j^{\rmth}$ Betti number of the variety,
$b^j(X)={\rm dim~H}^j_{\rm dR}(X)$. The rationality of the zeta
function was first shown by Dwork \cite{d60} by adelic methods.

The resulting building blocks given by the polynomials
$\cP_p^j(t)$ associated to the full cohomology group $H^j(X)$ are
not useful in the present context, leading to L$-$series whose
Mellin transforms in general cannot directly be identified with
string theoretic modular forms of the type considered in
\cite{su02, ls04, rs05, rs06, kls08} and in the present paper. The
idea instead is to decompose these objects further, which leads to
the factorization of the polynomials $\cP_p^i(t)$.

{\bf 2)} The most difficult part of the Weil conjectures is
concerned with the nature of the factorization of the polynomials
 \beq
 \cP^j_p(t)
 = \prod_{i=1}^{b^j} \left(1-\g^j_i(p) t\right).
 \lleq{weil-grothendieck-factorization}
 Experience with Jacobi
 and Gauss sums in the context of diagonal weighted
projective varieties indicates that the inverse eigenvalues
$\g^j_i(p)$ are algebraic integers that satisfy the Riemann
hypothesis
  \beq
     |\g^j_i(p)| = p^{j/2},~~~~\forall i.
  \eeq
  It is this part of the Weil conjectures which resisted the longest,
and was finally proved by Deligne \cite{d74}.

{\bf 3)} Given the algebraic nature of the inverse roots of the
 polynomials $\cP_p^j(t)$ one can consider the field
 $K=\mathQ(\{\g^j_i|j=0,...,2d-1,i=1,...,b^j\})$.
 The field $K$ is separable and therefore one can consider orbits
within the cohomology with respect to the embedding monomorphisms.
For $n-$dimensional Calabi-Yau varieties one can, in particular,
consider the orbits $\cO_{\Om}$ associated to the holomorphic
$n-$forms $\Om \in H^{n,0}(X)$, while for special Fano manifolds
of charge $Q$ one can consider $\Om \in H^{n-(Q-1),(Q-1)}(X)$. The
orbits of these forms generated by the embeddings of the field $K$
lead to a projection $p_{\Om}$ on the intermediate cohomology,
leading to the Grothendieck motive  $M_{\Om}= (X,p_{\Om},Q)$. This
will be called the $\Om-$motive of the variety $X$ in both, the
Calabi-Yau case and the more general case of special Fano
varieties.

{\bf 4)} Given the $\Om-$motive $M_{\Om}$ one can combine the
local factors of the zeta functions, leading to the motivic
L-function of the variety
 \beq
 L_{\Om}(X,s) := L(M_{\Om},s) =
 \prod_p \frac{1}{\cP^{\Om}_p(p^{-s})} ,
 \eeq
 where $\cP_p^{\Om}(t)$ are the polynomials described by the
 orbit of $\Om \in H^{d-(Q-1),(Q-1)}(X)$. Denoting the Weil number
 corresponding to $\Om$ in (\ref{weil-grothendieck-factorization})
 by $\g_{\Om}$ the polynomial can be expressed as
 \beq
  \cP_p^{\Om}(t) = \prod_{\si \in \rmHom(K,\mathC)}
     (1- \si(\g_{\Om})t).
  \eeq
The $\si-$orbits defined via the embedding monomorphisms
 define traces of the number field $K$, which implies that the
 corresponding $\Om-$motives have L-functions
 $L_{\Om}(X,s)$ with integral coefficients.

\subsection{String theoretic modularity and automorphy}

With the above structures in place we can ask in full generality
for any Calabi-Yau variety the following

 {\bf Questions.} \hfill \break
 When is the L-function $L_{\Om}(X,s)$ of the $\Om-$motive
 $M_{\Om}$ of a Calabi-Yau variety modular? Further, if it is
 modular, can $L_{\Om}(X,s)$ be expressed in terms of string theoretic forms
 associated to Kac-Moody algebras?

Modularity of $L_\Om(X,s)$ here is understood to include the usual
linear operations on modular forms.

More generally this question can be raised for the class of
motives associated to Fano varieties of special type. It has been
shown in refs. \cite{su02, ls04, rs05,rs06} that the answer to
this question is affirmative at least sometimes in lower
dimensions, and generalizations to higher dimensions will be
established below. In such cases the L-function can be viewed as a
map that takes motives and turns them into conformal field
theoretic objects. This framework therefore leads to the following
picture: \hfill \break
 {\bf Conjecture:} \hfill \break
  The L-function provides a map from the category of Calabi-Yau motives
  (more general special Fano type motives)
  to the category of $N=2$ supersymmetric conformal field theories.

 The question and the conjecture can be raised in the more general
 context of automorphic representations. In this case the
 Langlands program leads to the expectation that every motive is
 automorphic, and at this level of speculation the question
 becomes whether the resulting automorphic forms have a
 string theoretic interpretation.

\subsection{Grothendieck motives for Brieskorn-Pham varieties}

In ref. \cite{rs06} the notion of an $\Om-$motive was introduced
because it leads to geometric theta series with coefficients that
are rational integers. This allows in principle to compare these
series with those obtained from (modified) characters in the
conformal field theory. In a first approximation the $\Om-$motive
can roughly be viewed via its realization in cohomology, given by
the Galois orbit of the holomorphic forms on Calabi-Yau manifolds
as well as Fano varieties of the special type considered in
\cite{rs92, cdp93, rs94}. A more intrinsically geometric
perspective is provided by Grothendieck's notion of a motive. The
basic strategy to construct Grothendieck motives from
$\Om-$motives is to construct algebraic correspondences via
projectors that are associated to Galois orbits. These projectors
define algebraic cycles which then can be used to define the
correspondences. In the context of Fermat varieties such a
transition has been constructed explicitly by Shioda \cite{s87}
(see also \cite{gy95, y06, ky08}).

Consider a diagonal hypersurface $X_n^d$ of degree $d$ and
dimension $n$ in a weighted projective space with weights
$(k_0,...,k_{n+1})\in \mathN^{n+2}$. With $d_i = d/k_i$ one can
consider the group $G^d_n = \prod_i (\mu_{d_i})$, where
$\mu_{d_i}$ is the cyclic group with generator $\xi_{d_i}=e^{2\pi
i/d_i}$. This group acts on the projective space as
 \beq
  gz=(\xi_{d_0}^{a_0}z_0,...,\xi_{d_{n+1}}^{a_{n+1}}z_{n+1})
 \eeq
 for a vector $a=(a_0,...,a_{n+1}) \in \mathZ^{n+2}$.
 Motives of Fermat type can be defined via projectors that are
 associated to the characters of the symmetry group. Denote the
 dual group of $G_n^d$ by $\Gwhat_n^d$ and associate to $a\in
 \Gwhat_n^d$ a projector $p_a$ as
 \beq
 p_a = \frac{1}{|G_s^d|} \sum_{g\in G_n^d} a(g)^{-1}g,
 \eeq
 where the character defined by $a$ is given by
  \beq
  a(g) = \prod_{i=0}^{n+1} \xi_{d_i}^{a_i}.
  \eeq
 Combining the projectors $p_a$ within a
 $\rmGal(\mathQ(\mu_d)/\mathQ)-$orbit then leads to
 \beq
  p_{\cO} := \sum_{a \in \cO} p_a.
 \eeq

 The projectors $p_{\cO}$ can now be regarded as algebraic
 cycles on $X\times X$ with rational coefficients by considering
 their associated graphs. By abuse of notation
 effective motives of Fermat type can therefore be defined as
 objects that are determined by the Galois orbits $\cO$ as
  $(X_s^n, p_{\cO})$. Including Tate twists then leads to
  Grothendieck motives
 \beq
   M_{\cO}:= (X, p_{\cO}, m).
 \eeq

 \subsection{$\Om-$motives for CY and SF BP hypersurfaces}

 For varieties of Calabi-Yau and special Fano type there exists a
 particularly important cohomology group which can be
 parametrized explicitly. For Calabi-Yau manifolds of complex
 dimension $n$ these forms take values in $H^{n,0}(X)$, while for
 special Fano varieties of complex dimension $n$ and charge $Q$ one has more generally
 $\Om \in H^{n-(Q-1),(Q-1)}(X)$, where $Q\in \mathN$.
 The $\Om-$motive associated to the Galois orbit of $\Om$ of a variety
 of charge $Q$ will be denoted by
 \beq
  M_{\Om}^Q = (X,p_{\Om},Q-1).
 \eeq
 For $Q=1$ one recovers the Calabi-Yau case, which will be denoted
 by $M_{\Om}=M_{\Om}^1$.

\subsection{Lower weight motives from higher dimensional
varieties}

The $\Om-$motive is not the only motive of Calabi-Yau and special
Fano varieties that can lead to interesting modular forms.
Instead, one can construct Grothendieck motives $M_\cO(X)$
associated to Galois orbits $\cO$ that do not arise from the
$\Om-$form, but instead come from motives represented by subgroups
of the remainder of the intermediate cohomology group. The
associated L-series $L(M_\cO(X),s)$ may, or may not lead to
interesting modular forms. In the case they do in the examples
below, the resulting modular forms can be described as determined
by Tate twists of forms
 $$
 L(M_\cO(X),s) = L(f_\cO,s-1),
 $$
 where $f_\cO(q)$ is a modular associated to the motive
 $M_\cO(X)$. In general one would expect that these lower weight
 motives lead to Tate twists of automorphic forms of lower dimensional
 varieties that are embedded in the higher-dimensional manifolds.

\vskip .2truein

\section{L-functions for Calabi-Yau varieties of dimension 2,3, and 4}

In this paper the focus is on Calabi-Yau manifolds in complex
dimensions two, three and four. String theoretic examples of rank
two and dimension one and two are considered in
\cite{su02,ls04,rs05,rs06}. Combining these examples with the
varieties discussed in the present paper covers all physically
interesting dimensions in string theory, M-theory, and F-theory.

A general, non-toroidal, Calabi-Yau 2-fold $X_2$ is a K3 surface,
i.e with Betti numbers $b^1=0=b^3$, leading to the zeta function
 \beq
  Z(X_2/{\mathbb F}_p, t)= \frac{1}{(1-t)\cP^2_p(t)(1-p^2t)}
 \eeq
 with $\rmdeg(\cP^2_p(t)) = b^2 = 22$.
Expanding this rational form via
 \beq
 \cP^2_p(t) = \sum_{i=0}^{22} \beta^2_i(p) t^i
 \eeq
 leads to
 \beq
 \beta^2_1(p) = 1+p^2-N_{1,p}.
 \eeq
The only interesting L-function associated to a K3 surface
therefore is associated to its second cohomology group, in
particular its $\Om-$motivic piece, leading to $L_{\Om}(X_2,s)$.

Calabi-Yau threefolds with finite fundamental group lead to zeta
functions of the form
 \beq
 Z(X/{\mathbb F}_p, t)= \frac{\cP^3_p(t)}{(1-t)\cP^2_p(t)\cP^4_p(t)(1-p^3t)}
 \eeq
 with
 \begin{eqnarray}
   \rmdeg(\cP^3_p(t)) &=& 2+2h^{(2,1)} \nonumber \\
   \rmdeg(\cP^2_p(t)) &=& h^{(1,1)} \nonumber
\end{eqnarray}
This follows from the fact that for non-toroidal Calabi-Yau
threefolds we have $b^1=0$. For Calabi-Yau threefolds with
$h^{1,1}=1$ the zeta function reduces to
 \beq
 Z(X/{\mathbb F}_p, t)=
 \frac{\cP^3_p(t)}{(1-t)(1-pt)(1-p^2t)(1-p^3t)}
 \eeq
  and therefore becomes particularly simple. The coefficients
$\beta^3_i(p)$ of the polynomial
 \beq
 \cP^3_p(t) = \sum_{j=0}^{b^3(X)} \beta^3_i(p)t^i
 \eeq
are related to the cardinalities of the variety via the expansion
 \beq
 Z(X/\mathF_p,t)
 =  1 + N_{1,p}t + \frac{1}{2}(N_{1,p}^2 + N_{2,p})t^2 +
      \left(\frac{1}{3}N_{3,p} +\frac{1}{2}N_{1,p}N_{2,p} +
          \frac{1}{6}N_{1,p}^3\right)t^3 + \cO(t^4)
 \eeq
  as
 \bea
 \beta^3_1(p) &=& N_{1,p} - (1+p+p^2+p^3) \nn \\
 \beta^3_2(p) &=& \frac{1}{2}(N_{1,p}^2+N_{2,p}) -
  N_{1,p}(1+p+p^2+p^3) \nn \\ & &~+(1+p+p^2+p^3)^2 + (1+p+p^2) +
 p^2(1+p) + p^4(1+p+p^2)
 \eea
 etc.
This procedure is useful because the knowledge of a finite number
of terms in the L-function determines it uniquely.

In dimension four the cohomology of Calabi-Yau varieties is more
complicated, leading to zeta functions
 \beq
  Z(X_4/\mathF_p,t)
  = \frac{\cP_p^3(t)\cP^5(t)}{(1-t)\cP_p^2(t)\cP_p^4(t)\cP_p^6(t)(1-p^4t)}
 \eeq
 for non-toroidal spaces, but the procedure is the same as above. For smooth
 hypersurfaces the cohomology groups except of degree given by the dimension are either
 trivial or inherited from the ambient space, leading to the
intermediate L-function
 \beq
  L(X,s) = \prod_p \frac{1}{\cP^4_p(p^{-s})}
 \eeq
 as the only nontrivial factor.

\section{L-functions via Jacobi sums}

For the class of hypersurfaces of Brieskorn-Pham type it is
possible to gain insight into the precise structure of the
L-function by using a result of Weil \cite{w49} which expresses
the cardinalities of the variety in terms of Jacobi sums of finite
fields. In this context there the L-function of the $\Om-$motive
of such weighted Fermat hypersurfaces can be made explicit.

For any degree vector $\un=(n_0,...,n_{s+1})$ and for any prime
$p$ define the numbers $d_i=(n_i,p-1)$ and the set
 \beq
 \cA_s^{p,\un} = \left\{(\a_0,...,\a_{s+1}) \in \mathQ^{s+2}~|~
 0<\a_i<1, d_i\a_i =0~(\rmmod~1), \sum_i \a_i =0
~(\rmmod~1)\right\}.
 \eeq

{\bf Theorem 5.}~ {\it The number of solutions of the smooth
projective variety}
 \beq
 X_s = \left\{(z_0:z_1:\cdots :z_{s+1}) \in
\mathP_{s+1}~|~\sum_{i=0}^{s+1} b_iz_i^{n_i}=0 \right\} \subset
\mathP_{s+1}
 \eeq
{\it over the finite field $\mathF_p$ is given by}
 \beq
    N_p(X_s) = 1 + p + p ^2 + \cdots + p^s +
  \sum_{\a \in \cA_s^{p,n}} j_p(\a) \prod \bchi_{\a_i}(a_i),
  \eeq
  {\it where}
   \beq
   j_p(\a) = \frac{1}{p-1} \sum_{\stackrel{u_i
\in \IF_q}{ u_0 + \cdots + u_s=0}} \chi_{\a_0}(u_0)\cdots
\chi_{\a_s}(u_s).
 \eeq

With these Jacobi sums $j_q(\a)$ one defines the polynomials
 \beq
 \cP_p^s(t) = (1-p^{s/2}t)^{|s|} \prod_{\a \in \cA_s^n}
 \left(1-(-1)^sj_{p^f}(\a)\prod_i\bchi_{\a_i}(b^i) t^f\right)^{1/f}
 \eeq
 and the associated L-function
  \beq
  L^{(j)}(X,s) =
\prod_p \frac{1}{\cP^j_p(p^{-s})}.
\eeq
 Here $|s|=1$ if $s$ is even and $|s|=0$ if $s$ is odd.

A slight modification of this result is useful even in the case of
smooth weighted projective varieties because it can be used to
compute the factor of the zeta function coming from the invariant
part of the cohomology, when viewing these spaces as quotient
varieties of projective spaces.

\vskip .1truein

The Jacobi-sum formulation allows to write the L-function of the
$\Om-$motive $M_{\Om}$ of weighted Fermat hypersurfaces in a more
explicit way. Define the vector
 $\a_{\Om} = \left(\frac{k_0}{d},..., \frac{k_{s+1}}{d}\right)$
 corresponding
to the holomorphic $s-$form, and denote its Galois orbit by
$\cO_{\Om} \subset A_s^{\un}$. Then
 \beq
 L_{\Om}(X,s) ~=~ \prod_p \prod_{\a \in \cO_{\Om}}
  \left(1-(-1)^s j_{p^f}(\a)p^{-fs}\right)^{-1/f}.
 \eeq

\vskip .2truein

\section{Modularity results for rank 2 motives in dimension one
and two}

The simplest possible framework in which this question can be
raised is for toroidal compactifications, in particular
Brieskorn-Pham curves $E^d$ of degree $d$ embedded in the weighted
projective plane. The elliptic curves $E^d$ are defined over the
rational number $\mathQ$, and therefore modular in lieu of the
Shimura-Taniyama conjecture, proven in complete generality in ref.
\cite{bcdt01}, based on Wiles' breakthrough results \cite{w95}.
This theorem says that any elliptic curve over the rational
numbers is modular in the sense that the inverse Mellin transform
of the Hasse-Weil L-function is a modular form of weight two for
some congruent subgroup $\Gamma_0(N)$. This raises the question
whether the modular forms derived from these Brieskorn-Pham curves
are related in some way to the characters of the conjectured
underlying conformal field theory models.

The conformal field theory on the string worldsheet is fairly
involved, and a priori there are a number of different modular
forms that could play a role in the geometric construction of the
varieties. The first string theoretic modularity result showed
that the modular form associated to the cubic Fermat curve $E^3
\subset \mathP_2$ factors into a product of SU(2)$-$modular forms
that arise from the characters of the underlying world sheet
\cite{su02}. More precisely, the worldsheet forms that encode the
structure of the compact spacetime geometry are the Hecke
indefinite theta series associated to Kac-Moody theoretic string
functions introduced by Kac and Peterson. For the remaining two
elliptic weighted Fermat curves this relation requires a
modification involving a twist character that is physically
motivated by the number field generated by the quantum dimensions
of the string model \cite{ls04, rs05}. The explicit structure of
these one-dimensional results is described in Section 6. The
modular forms that emerge from the weighted Fermat curves provide
a string theoretic interpretation of the Hasse-Weil L-function of
the exactly solvable Gepner models at central charge $c=3$. Ref.
\cite{rs05} also identifies the criteria that lead to the
derivation of these elliptic curves from the conformal field
theory itself, with no a priori input from the geometry.

This Section briefly summarizes the results obtained in
\cite{su02,ls04,rs05,rs06}, several of which appear as building
blocks for the new examples considered in the remainder of this
paper.

Explicitly, the class of elliptic Brieskorn-Pham curves is given
by
 \bea
 E^3 &=& \left\{(z_0:z_1:z_2) \in \mathP_2~|~z_0^3+z_1^3+z_2^3=0
            \right\} \nn \\
 E^4 &=& \left\{(z_0:z_1:z_2) \in \mathP_{(1,1,2)}~|~z_0^4+z_1^4+z_2^2=0
          \right\} \nn \\
 E^6 &=& \left\{(z_0:z_1:z_2) \in \mathP_{(1,2,3)}~|~z_0^6+z_1^3+z_2^2=0
 \right\}.
 \llea{elliptic-examples}
 The modular forms associated to these curves are cusp forms of
 weight two with respect to congruence groups of level $N$
 $\G_0(N) \subset \rmSL(2,\mathZ)$ defined by
 \beq
  \G_0(N) = \left\{\left(\matrix{a&b\cr c&d\cr}\right)
   \in \rmSL_2(\ZZ) ~{\Big |}~\left(\matrix{a&b\cr c&d\cr}\right)
   \sim \left(\matrix{*&*\cr 0&*\cr}\right) ~(\rmmod~N)\right\}.
 \eeq
 Recalling the string theoretic Hecke indefinite theta series
 $\Theta^k_{\ell,m}$ defined in Section 2  in terms of the
  Kac-Peterson string functions $c^k_{\ell,m}(\tau)$, the
 geometric modular forms roughly decompose as
 \beq
 f(E^d,q) = \prod_i \Theta^{k_i}_{\ell_i,m_i}(q^{a_i})\otimes
 \chi_d,
 \eeq
 where $\chi_d(\cdot) = \left(\frac{d}{\cdot}\right)$ is the Legendre
 symbol. More precisely, given the worldsheet
 theta series $\Theta^1_{1,1}(\tau) =\eta^2(\tau)$ and
 $\Theta^2_{1,1}(\tau)=\eta(\tau)\eta(2\tau)$ the factorization
 takes the following form \cite{rs05}.

{\bf Theorem 6.} ~{\it The inverse Mellin transforms $f(E^d,q)$ of
the Hasse-Weil L-functions $L_{\rmHW}(E^d,s)$ of the curves $E^d,
i=3,4,6$ are modular forms $f(E^d,q) \in S_2(\Gamma_0(N))$, with
 $N = 27, 64, 144$ respectively. These cusp forms factor as}
 \bea
 f(E^3,q) &=& \Theta^1_{1,1}(q^3)\Theta^1_{1,1}(q^9) \nn \\
 f(E^4,q) &=& \Theta^2_{1,1}(q^4)^2\otimes \chi_2 \nn \\
 f(E^6,q) &=& \Theta^1_{1,1}(q^6)^2\otimes \chi_3.
 \llea{mod-forms-elliptic}

Consider the class of extremal K3 surfaces that can be constructed
as weighted Fermat varieties
  \bea
   X_2^4 &=& \left\{(z_0:\cdots :z_3)\in \mathP_3~{\Big
   |}~ z_0^4 +z_1^4+z_2^4+z_3^4 = 0 \right\}, \nn \\
   X_2^{6\rmA} &=& \left\{(z_0:\cdots :z_3)\in \mathP_{(1,1,1,3)}~{\Big
   |}~ z_0^6+z_1^6 + z_2^6+z_3^2=0 \right\}, \nn \\
   X_2^{6\rmB} &=& \left\{(z_0:\cdots :z_3)\in \mathP_{(1,1,2,2)}~{\Big
   |}~ z_0^6+z_1^6 + z_2^3+z_3^3=0 \right\}.
 \llea{extremal-k3s}
These surfaces have motives that are modular and admit a string
theoretic interpretation. Denote by $M(X) \subset H^2(X_2^d)$ the
cohomological realization of a motive $M$, with $M_\Om(X)$ the
motive associated to the holomorphic 2-form, and let $L_\Om(X,s) =
L(M_\Om(X),s)$ be the associated L-series, with $f_\Om(X,q)$
denoting the inverse Mellin transform of $L_\Om(X,s)$. The
following result shows that the modular forms determined by the
$\Om-$motives of these extremal K3 surfaces are determined by the
string theoretic modular forms determined in Theorem 6.

 {\bf Theorem 7.}~ {\it Let $M_{\Om} \subset H^2(X_2^d)$ be the irreducible
 representation
 of $\rmGal(\mathQ(\mu_d)/\mathQ)$ associated to the holomorphic 2$-$form
 $\Om \in H^{2,0}(X_2^d)$ of the K3 surface $X_2^d$, where $d=4,6\rmA, 6\rmB$.
 Then the $q-$series $f_{\Om}(X_2^d,q)$ of the L-functions $L_{\Om}(X_2^d,s)$
 are modular forms given by}
  \bea
   f_{\Om}(X_2^4,q) &=& \eta^6(q^4) \nn \\
   f_{\Om}(X_2^{6\rmA},q) &=& \vartheta(q^3) \eta^2(q^3)\eta^2(q^9) \nn \\
   f_{\Om}(X_2^{6\rmB},q) &=& \eta^3(q^2)\eta^3(q^6) \otimes \chi_3.
  \llea{extr-k3mod}
 {\it These functions are cusp forms of weight three with respect
 to $\Gamma_0(N)$ with levels $16, 27$ and 48, respectively.
 For $X_2^4$ and $X_2^{6\rmA}$ the L-functions can be written as}
 \beq
   L_{\Om}(X_2^d,s) = L(\psi_d^2, s),
 \eeq
 {\it  where $\psi_d$ are algebraic Hecke characters associated to
   cusp forms $f_d(q)$ of weight two and levels 64 and 27,
   respectively, given by the elliptic forms}
   \bea
    f_4(\tau) &=& f(E^4,q) = \eta^2(q^4)\eta^2(q^8)\otimes \chi_2 \nn \\
    f_{6\rmA}(\tau) &=& f(E^3,q) = \eta^2(q^3)\eta^2(q^9).
   \eea
  {\it For $X_2^{6\rmB}$ the L-series is given by
   $L_{\Om}(X_2^{6\rmB},s) = L(\psi_{144}^2\otimes \chi_3,s)$, leading to
   the cusp form of level 144}
   \beq
    f_{6\rmB}(\tau) = f(E^6,q) = \eta^4(q^6)\otimes \chi_3.
   \eeq

These results will enter in the discussion below of higher
dimensional varieties.

 \vskip .2truein

\section{A nonextremal K3 surface $X_2^{12} \subset \mathP_{(2,3,3,4)}$}

String theoretic modularity of the class of extremal K3 surfaces
of Brieskorn-Pham type has been established in \cite{rs06}.
Extremal K3 surfaces are characterized by the fact that their
Picard number is maximal, i.e. $\rho =20$.

A nonextremal example of a K3 surface is defined as the
Brieskorn-Pham hypersurface (\ref{k3deg12}) of degree twelve in
the weighted projective space $\mathP_{(2,3,3,4)}$. The Galois
group of the cyclotomic field $\mathQ(\mu_{12})$ has order four,
hence the $\Om-$motive has rank four. The four Jacobi sums which
parametrize this motive are given by
 \beq
  j_p(\si \a_{\Om}),~~~~\si \in \rmGal(\mathQ(\mu_{12})/\mathQ)
 \eeq
 i.e. $j_p(\a)$ with
 \beq
  \a \in \left\{\left(\frac{1}{6},\frac{1}{4},\frac{1}{4},\frac{1}{3}\right),
  ~~\left(\frac{5}{6},\frac{1}{4},\frac{1}{4},\frac{2}{3}\right)\right\}
 \eeq
 and their conjugates $\balpha = {\bf 1} - \a$, where ${\bf 1}$ denotes the
 unit vector. The values of the independent Jacobi sums are collected for
 low $p^f$ in Table 1.
 \begin{center}
\begin{tabular}{c| r r}
 $p^f$
   &$j_{p^f}\left(\frac{1}{6},\frac{1}{4},\frac{1}{4},\frac{1}{3}\right)$
   &$j_{p^f}\left(\frac{5}{6},\frac{1}{4},\frac{1}{4},\frac{2}{3}\right)$
                    \\

\hline

 $13$
   &$-3 + 4\sqrt{3} + \left(2+6\sqrt{3}\right)i$
   &$-3 - 4\sqrt{3} + \left(2-6\sqrt{3}\right)i$
 \tabroom \\

 $25$
   &$15 - 20i$
   &$15 - 20i$
   \tabroom \\

 $37$
   &$-5 - 12\sqrt{3} + \left(30-2\sqrt{3}\right)i$
   &$-5 + 12\sqrt{3} + \left(30+2\sqrt{3}\right)i$
   \tabroom \\

 $49$
   &$-7 - 28\sqrt{3}i$
   &$-7 + 28\sqrt{3}i$
   \tabroom \\

 $61$
  &$35 - 12\sqrt{3} - \left(42+10\sqrt{3}\right)i$
  &$35 + 12\sqrt{3} - \left(42-10\sqrt{3}\right)i$
        \tabroom \\

 $73$
  &$15 - 32\sqrt{3} + \left(40+12\sqrt{3}\right)i$
  &$15 + 32\sqrt{3} + \left(40-12\sqrt{3}\right)i$
     \tabroom \\
 \hline
\end{tabular}
\end{center}

\centerline{{\bf Table 1.}~{\it Jacobi sums for the K3 surface
$X_2^{12} \subset \mathP_{(2,3,3,4)}$.}}

Using these results leads to the
 L-function of the $\Om-$motive
 \beq
 L_{\Om}(X_2^{12},s)
 \doteq 1 - \frac{12}{13^s} + \frac{30}{25^s}
          - \frac{20}{37^s} - \frac{14}{49^s}
         + \frac{140}{61^s} +
          \frac{60}{73^s} +  \cdots
 \eeq

Insight into the structure of this L-function can be obtained by
noting that the surface $X_2^{12}$ can be constructed via the
twist map \cite{hs96,hs99} (see also \cite{v93,b97}) by
considering
 \beq
 \Phi: \mathP_{(2,1,1)} \times \mathP_{(3,1,2)} ~~ \lra ~~
  \mathP_{(3,3,2,4)}
 \eeq
 defined by
 \beq
 ((x_0,x_1,x_2),(y_0,y_1,y_2)) ~~\mapsto ~~
 \left(y_0^{1/3}x_1,y_0^{2/3}x_2, x_0^{1/2}y_1,
 x_0^{1/2}y_2\right),
 \eeq
 which on the product $E^4_{-}\times E^6$ leads to the
 K3 surface of degree twelve $X_2^{12}$.

The coefficients of $L(E^4_-,s) = L(E^4,s) = \sum_n
a_n(E^4)n^{-s}$ and $L(E^6,s) = \sum_n b_n(E^6)n^{-s}$ of the
Hasse-Weil L-functions of the elliptic curves $E^4$ and $E^6$,
respectively, can be obtained by expanding the results of Theorem
6.
 Multiplying the coefficients $a_p(E^4)$ and $b_p(E^6)$ leads
 \beq
  a_p(E^4) b_p(E^6) = c_p(X_2^{12}).
 \eeq
For low primes the results are collected in Table 2.

 \begin{center}
 \begin{tabular}{l| r r r r r r}

 $p^f$           &$13$   &25     &$37$    &49    &$61$   &$73$
  \\
\hline

$a_{p^f}(E^4)$   &$-6$   &$-1$   &$2$     &$-7$
                          &$10$            &$-6$   \tabroom \\

$b_{p^f}(E^6)$   &$\phantom{+}2$
                         &$-5$   &$-10$   &9
                           &$14$            &$-10$
                                        \tabroom \\

\hline

 $c_{p^f}(X_2^{12})$  &$-12$  &\phantom{+}30   &$-20$  &$-14$  &140
                          &\phantom{+}60    \tabroom \\

\hline
\end{tabular}
\end{center}

\centerline{{\bf Table 2.}~{\it Coefficient comparison of the
surface $X_2^{12}$ and the curves $E_-^4$ and $E^6$.}}

The Mellin transform of the L-function of both building blocks
$E^4$ and $E^6$ of the surface $X_2^{12}$ are given in terms of
string theoretic theta functions as described in Theorem 6.
 The modular forms of $E^4$ and $E^6$ are both of complex
 multiplication type, leading to Hecke interpretation of the
 L-series in terms of Gr\"o\ss encharaktere. They are both twists
 by Legendre symbols of characters $\psi_{32}$ and $\psi_{36}$, as
 described in \cite{rs05}, leading to
 \bea
  L(E^4,s) &=& L(\psi_{32}\otimes \chi_2,s) \nn \\
  L(E^6,s) &=& L(\psi_{36}\otimes \chi_2,s),
 \eea
 where the characters $\psi_{32}$ and $\psi_{36}$ are associated to
  the Gauss field $\mathQ(\sqrt{-1})$ and the Eisenstein field $\mathQ(\sqrt{-3})$
 respectively, as described in \S 2.

 The string interpretation of the Hasse-Weil L-series of $E^4$ and $E^6$ described in
 Theorem 6 therefore induces a string interpretation of the modular blocks of
 the K3 surface $X_2^{12}$.  The CM property of these forms will
 allow a systematic discussion in Section 12 of the reverse construction of emergent
 space from the modular forms of the worldsheet field theory.

 The factorization of the coefficients $c_p(X_2^{12})$ suggests that
 the rank four $\Om-$motive $M_{\Om}(X_2^{12})$ is the tensor
 product of the elliptic motives of $E^4$ and $E^6$, but a priori leaves
 open the precise nature of the L-function product. It turns out
 that the correct version is the modified Rankin-Selberg product
 considered in Section 2, applied to the case of two modular forms
 of weight two
 $$
 L_{\Om}(X_2^{12},s) = L(M_{f_{64}}\otimes M_{f_{144}},s),
 $$
 where $M_{f_{N_i}}$ are the elliptic motives of $f_{N_i}$.

\vskip .2truein

\section{Modular motives of the Calabi-Yau threefold
 $X_3^6 \subset \mathP_{(1,1,1,1,2)}$}

In this and the following section modularity is established for
the $\Om-$motives of two Calabi-Yau threefolds. The two varieties
considered lead to motives of ranks two and four.

\subsection{The modular $\Om-$motive of $X_3^6$}

Consider the Calabi-Yau variety $X_3^6$ defined as the double
cover branched over the degree six Fermat surface in projective
 threespace $\mathP_3$. This manifold can be viewed as a smooth
 degree six hypersurface of Brieskorn-Pham type in the weighted projective
 fourspace $\mathP_{(1,1,1,1,2)}$ as in (\ref{modular-3folds}).
 The Galois group of $\mathQ(\mu_6)$ has order two, hence
 the motive $M_{\Om}$ has rank two. The values of the
  relevant Jacobi sum $j_{p^f}(\a_{\Om})$ with
  $\a_{\Om}= \left(\frac{1}{6},\frac{1}{6},\frac{1}{6},
  \frac{1}{6},\frac{1}{3}\right)$ are collected in Table 3.

\begin{small}
 \begin{center}
\begin{tabular}{l| c c c c c c}

 $p^f$         &7    &13   &19  &25 &31  &37    \tabroom \\

\hline

 & & & & & & \\

$j_{p^f}(\a_{\Om})$   &$-\frac{17}{2} - \frac{19}{2}\sqrt{3}i$
                  &$-\frac{89}{2}-\frac{17}{2}\sqrt{3}i$
                  &$-\frac{107}{2} + \frac{73}{2}\sqrt{3}i$
                  &125
                  &$-154 -45\sqrt{3}i$
                  &$\frac{433}{2} - \frac{71}{2}\sqrt{3}i$
                    \\

                  & & & & & & \\

\hline
\end{tabular}
\end{center}
\end{small}

 \centerline{{\bf Table 3.}~{\it Values for the Jacobi sum of
the $\Om-$motive of $X_3^6 \subset \mathP_{(1,1,1,1,2)}$.}}

These Jacobi sums and their conjugates lead to the L-series of the
motive of $X_3^6$
 \beq
  L_{\Om}(X_3^6,s) \doteq 1 + \frac{17}{7^s} + \frac{89}{13^s}
                        + \frac{107}{19^s} - \frac{125}{25^s}
                        + \frac{308}{31^s} - \frac{433}{37^s} + \cdots
 \eeq
 The inverse Mellin transform of this L-series turns out to
 describe a
 modular form of weight $w=4$ and level $N=108$. This form
 $f_\Om \in S_4(\G_0(108))$ cannot be written as a
 product or a quotient of Dedekind eta-functions \cite{dkm85, ym96},
 but it admits complex multiplication and $L_{\Om}(X_3^6,s)$ can be written
 as the Hecke L-series of a twisted Gr\"o\ss encharakter associated to the
 complex multiplication field $K=\mathQ(\sqrt{-3}$.
 Characters associated to $K$ were
 considered in \cite{rs05, rs06} in the context of the elliptic
 Brieskorn-Pham curve $E^3 \subset \mathP_2$ and $E^6 \subset
 \mathP_{(1,2,3)}$ as well as modular K3 surfaces. The Hasse-Weil
 L-series $L(E^3,s)$ is a Hecke series for the
 character $\psi_{27}$ considered in \S 2 which can be written as
 \cite{su02}
  \beq
  L(E^3,s)=L(\psi_{27},s)=L(\Theta^1_{1,1}(q^3)\Theta^1_{1,1}(q^9),s)
 \lleq{string-e3}
 The character $\psi_{27}$ turns out to be the fundamental
 building block of the L-series of the $\Om-$motive of $X_3^6$.

 For higher dimensional varieties it is possible to generalize such
 relations by considering powers
 of the Hecke character in order to obtain higher weight modular
 forms, as described above in Hecke's theorem. In the present case
 $\psi_{27}^3$ leads to a modular form which can be written in terms
 of the Dedekind eta function $\eta(q)$  as
 $\eta^8(q^3) \in S_4(\G_0(9))$, which also appears as a motivic form
 in a number of geometries
 different from $X_3^6$. In order to obtain the motivic
 L-series $L_\Om(X_3^6,s)$ computed above it is necessary to introduce a twist
 character. This can be chosen to be the cubic residue power
 symbol, denoted here by
  \beq
   \chi_2^{(3)}(p) := \left(\frac{2}{\pfrak}\right)_3,
  \eeq
  where $\pfrak|p$ and the congruence ideal is chosen to be
  $\mfrak=(3)$. With this character the Hecke interpretation of
  the motivic L-function of $X_3^6$ takes the form
  \beq
   L_{\Om}(X_3^6,s) = L(\psi_{27}^3 \otimes (\chi_2^{(3)})^2,s).
  \eeq
The inverse Mellin transform $f_{\Om}(X_3^6,q)$ of this L-series
is a modular form of weight 4 and level 108
 \beq
  f_{\Om}(X_3^6,q) \doteq q + 17q^7 + 89q^{13} + 107q^{19} - 125q^{25}
             + 308q^{31}  - 433q^{37} + \cdots
  \lleq{f108}
 which is a cusp form, i.e. $f_{\Om}(X_3^6,q) \in S_4(\Gamma_0(108))$.
This modular form is therefore of complex multiplication type in
the sense of Ribet \cite{r77}. An explicit proof for the modular
form $f(\psi_{27},q) \in S_2(\G_0(27))$ can be found in
\cite{rs05}.

\subsection{Lower weight modular motives of $X_3^6$}

The degree six Calabi-Yau hypersurface $X_3^6$ provides an example
of the phenomenon noted in Section 3 that the intermediate
cohomology can lead to modular motives beyond the $\Om-$motive.
For $X_3^6$ the motives are of rank two, given by the Galois group
$\rmGal(\mathQ(\mu_6)/\mathQ)$ with certain multiplicities and
twists. Modulo these twists and multiplicities the group
$H^{2,1}(X)\oplus H^{1,2}(X)$ leads to three different types of
modular motives of weight two and rank two, denoted in the
following by $M_A \in \{M_\rmI, M_\rmII, M_\rmIII\}$.
 The L-series $L(M_A,s)$ that result from these motives have coefficients
 $\oa^A_p$ that are all divisible by the prime $p$. By introducing the
 twisted coefficients $a^A_p = \oa^A_p/p$, these L-series  lead
 to modular forms of weight two $f_A \in S_2(\G_0(N_A))$, where the
 level $N_A$ is determined by the motive $M_A$
 $$
 L(M_A(X_3^6),s) = L(f_A,s-1).
 $$
 The modular forms are of levels $N_A = 27, 144, 432$, the first
 two given by the curves $E^3, E^6$ described in Theorem 6
 \bea
  f_\rmI(q) &=& f(E^3,q) \nn \\
  f_\rmII(q) &=& f(E^6,q),
 \eea
 while the level $N_\rmIII=432$ form is given by
 $$
 f_\rmIII(q) \doteq q - 5q^7 - 7q^{13} + q^{19} + 4q^{31} + \cdots
 $$

It follows that the L-series $L(H^3(X_3^6),s)$ of the intermediate
cohomology group decomposes into modular pieces in the sense that
each factor arises from a modular form
 $$
 L(H^3(X_3^6),s) = L(f_\Om,s) \prod_i L(f_i\otimes
 \chi_i,s)^{a_i},
 $$
 where $a_i \in \mathN$, $f_\Om \in S_4(\G_0(108))$ is as determined above, the
 $f_i(q)$ are modular forms of weight two and levels $N_i =
 27,144,432$, and $\chi_i$ is a Legendre character (which can be trivial).

 The Jacobi sums corresponding to the motives $M_\rmI, M_\rmII, M_\rmII$
 are listed in Table 4, together with the level $N_A$ of the
 corresponding modular form $f_A \in S_2(\G_0(N_A))$.

\begin{center}
\begin{tabular}{c| c c}

Type     &Jacobi sum   &Level $N$  \tabroom \\

 \hline

 I  &$j_p\left(\frac{1}{3},\frac{1}{3},\frac{1}{3},\frac{1}{3},\frac{2}{3}\right)$
    &27                   \tabroom \\

 II  &$j_p\left(\frac{1}{6},\frac{1}{6},\frac{1}{2},\frac{5}{6},\frac{1}{3}\right)$
     &144                \tabroom \\

 III &$j_p\left(\frac{1}{6},\frac{1}{6},\frac{2}{3},\frac{2}{3},\frac{1}{3}\right)$
     &432              \tabroom \\

 \hline

\end{tabular}
\end{center}

\centerline{{\bf Table 4.} ~{\it The Jacobi sums of $X_3^6$ that
lead to non-isogenic modular motives.}}

The weight two modular forms $f_A$ that emerge from $X_3^6$ have a
natural geometric interpretation. The threefold $X_3^6$ contains
divisors given by degree six Fermat curves $C^6 \subset \mathP_2$
and $\oC^6 \subset \mathP_{(1,1,2)}$, obtained from the original
hypersurface via intersections with coordinate hyperplanes. These
curves are of genus ten and can be shown to decompose into ten
elliptic factors of three different types $E_\rmI=E^3,
E_\rmII=E^6$, and $E_\rmIII$ an elliptic curve of conductor 432.
Hence its L-function factors as
 $$
 L(C^6,s) = L(E_\rmI,s) L(E_\rmII,s)^6 L(E_\rmIII,s)^3,
 $$
 taking into account their multiplicities. The curve $\oC^6$ is of
 genus four and leads to the same modular forms, with different
 multiplicities.

\vskip .2truein

\section{The K3 fibration hypersurface $X_3^{12} \subset
\mathP_{(2,2,2,3,3)}$}

\subsection{The $\Om-$motive of $X_2^{12}$}

Consider the weighted Fermat hypersurface of degree twelve in
$\mathP_{(2,2,2,3,3)}$ given in eq. (\ref{modular-3folds}). The
Galois group of this variety is of order four, leading to an
$\Om-$motive of rank four. The Jacobi sums that parametrize this
motive are given by
  $j_p\left({\footnotesize
               \frac{1}{6},\frac{1}{6},\frac{1}{6},
                   \frac{1}{4},\frac{1}{4}}\right),
  j_p\left({\footnotesize \frac{1}{6},\frac{1}{6},\frac{1}{6},
                \frac{3}{4},\frac{3}{4}}\right)
$
 and their complex conjugates. The computation of these sums for
  low $p^f$ are collected in Table 5.
 \begin{center}
\begin{tabular}{c| r r}

 $p^f$ &$j_{p^f}\left({\footnotesize
               \frac{1}{6},\frac{1}{6},\frac{1}{6},
                   \frac{1}{4},\frac{1}{4}}\right)$
        &$j_{p^f}\left({\footnotesize \frac{1}{6},\frac{1}{6},\frac{1}{6},
                \frac{3}{4},\frac{3}{4}}\right)$    \tabroom \\

\hline

 13      &$-\frac{3}{2}+15\sqrt{3} +
              \left(1+\frac{45}{2}\sqrt{3}\right)i$
        &$-\frac{3}{2}-15\sqrt{3} - \left(1 - \frac{45}{2}\sqrt{3}\right)i$
                   \tabroom \\

 25     &$75-100i$
        &$75+100i$    \tabroom \\

 37     &$-\frac{47}{2}+99\sqrt{3} + \left(141 +
                             \frac{33}{2}\sqrt{3}\right)i$
       &$-\frac{47}{2}-99\sqrt{3} - \left(141 - \frac{33}{2}\sqrt{3}\right)i$
           \tabroom \\

 49     &$\frac{7}{2}(71+39\sqrt{3}i$
        &$\frac{7}{2}(71+39\sqrt{3}i$ \tabroom \\

 61     &$\frac{605}{2}+27\sqrt{3}-\left(363-\frac{45}{2}\sqrt{3}\right)i$
        &$\frac{605}{2}-27\sqrt{3}+\left(363+\frac{45}{2}\sqrt{3}\right)i$
                       \tabroom \\

73     &$-\frac{291}{2}+252\sqrt{3} - \left(388 +
            \frac{189}{2}\sqrt{3}\right)i$
       &$-\frac{291}{2}-252\sqrt{3} + \left(388 -
            \frac{189}{2}\sqrt{3}\right)i$
                              \tabroom \\

\hline
\end{tabular}
\end{center}

\centerline{{\bf Table 5.}~{\it Jacobi sums of the $\Om-$motive of
$X_3^{12}$.}}

Using these results leads to the expansion of the L-series
 \beq
  L_{\Om}(X_3^{12},s)
   \doteqdot 1 + \frac{6}{13^s} - \frac{150}{25^s} + \frac{94}{37^s}
       - \frac{497}{49^s} -\frac{1210}{61^s} + \frac{582}{73^s} + \cdots
 \eeq

The structure of the $\Om-$motivic L-series of $X_3^{12}$ can be
understood by noting that the threefold is a K3 fibration with
typical fiber $X_2^{6\rmA}$ given in (\ref{extremal-k3s}). The
interpretation of $L_{\Om}(X_3^{12},s)$ in terms of the fibration
is also useful because it makes the complex multiplication
structure of the associated modular form transparent. The
threefold can be constructed explicitly as the quotient of a
product of a torus $E$ and a K3 surface
 \beq
   X = E \times \rmK 3/\iota,
 \eeq
 where $\iota$ is an involution acting on the product.
More precisely, the elliptic curve is given by the weighted
 \beq
  E_-^4=\{x_0^2 - (x_1^4+x_2^4=0)\} \subset \mathP_{(2,1,1)}
 \eeq
 and the K3 surface is the generic fiber $X_2^{6\rmA}$.

 Applying the twist construction of \cite{hs96, hs99} gives first
 the map
 \beq
 \Phi:~ \mathP_{(2,1,1)} \times \mathP_{(3,1,1,1)} ~~\lra ~~
  \mathP_{(3,3,2,2,2)}
  \eeq
 defined by
  \beq
   ((x_0,x_1,x_2),(y_0,y_1,y_2,y_3)) ~\mapsto ~
   \left(y_0^{1/3}x_1, y_0^{1/3}x_2,
         x_0^{1/2}y_1, x_0^{1/2}y_2, x_0^{1/2}y_3\right).
  \eeq
  This map restricts on the product $E_{-}^4 \times X_2^{6\rmA}$ to the
  threefold $X_3^{12}$.

The fibration structure suggests that the L$-$function of the
threefold $X_3^{12}$ can be understood in terms of those of its
building blocks. The L-function of the K3 fiber of this threefold
was determined in Theorem 7 to be given by the Mellin transform of
the cusp form $f_{\Om}(X_2^{6\rmA},q)$ of weight $w=3$ and level
 $N=27$, and the L-function of the
 quartic curve is given by $f(E^4,q) \in S_2(\Gamma_0(64))$
 according to Theorem 6 \cite{rs05}.

 A comparison of the coefficients of the L-function of the $\Om-$motive of
 $X_3^{12}$ with those of its building blocks $E^4$ and $X_2^{6\rmA}$ should lead
 to a composite structure. Table 6 illustrates that this is indeed the case.
 The coefficients $a_p(E^4)$ arise from the L-series of the quartic
 $E^4$, while the surface expansion
 $b_p(X_2^{6\rmA})$  is that of (\ref{extr-k3mod}) in Theorem 7, which leads
 to the expansion
 \beq
 f(X_2^{\rm 6A},q)
 \doteq q - 13q^7 - q^{13} + 11q^{19} + 25q^{25} - 46q^{31} + 47q^{37}
  - 22q^{43} + 120q^{49} - 121q^{61} - 109q^{67} - 97 q^{73} + \cdots
 \eeq
  It follows that the products
 $a_p(E^4)b_p(X_2^{6\rmA})$ agree with the expansion coefficients
 $c_p(X_3^{12})$ of the threefold $X_3^{12}$.

 \begin{center}
 \begin{tabular}{c| c r r r r r r}

 $p^f$                &  &$13$     &25       &$37$  &49     &$61$     &$73$   \tabroom \\
 \hline

 $a_{p^f}(E^4)$         &  &$-6$     &$-1$     &2     &$-7$   &10       &$-6$    \tabroom \\

 $b_{p^f}(X_2^{6\rmA})$ &  &$-1$     &25       &47    &120    &$-121$   &$-97$    \tabroom \\

 \hline

 $c_{p^f}(X_3^{12})$
                    &  &6        &$-150$   &94    &$-994$  &$-1210$   &582   \tabroom \\

 \hline
 \end{tabular}
 \end{center}

\centerline{{\bf Table 6.}~{\it Coefficient comparison for the
threefold $X_3^{12}$.}}

The factorization of the $\Om-$motivic L-series
$L_{\Om}(X_3^{12},s)$ also shows that its building block have
complex multiplication. For the curve $E^4$ this was already
discussed above in the context of the K3 surface $X_2^{12}$
\cite{rs05}. For the K3 surface $X_2^{6\rmA}$ it was shown in
\cite{rs06} that the modular form of Theorem 7 is the Mellin
transform of the Hecke L-series of a Gr\"o\ss encharakter
 $\psi_{27}^2$ of the Eisenstein field $\mathQ(\sqrt{-3})$
 \beq
 L(X_2^{6\rmA},s) = L(\psi_{27}^2,s)
 \eeq
 where the character $\psi_{27}$ has been defined in Section 2.
 The motivic L-series of both building blocks of $X_3^{12}$
 therefore are of complex multiplication type.

\subsection{Lower weight modular forms of $X_3^{12}$}

Similar to the degree six threefold $X_3^6$ the cohomology
$H^{2,1}\oplus H^{1,2}$ of the degree twelve hypersurface
$X_3^{12}$ leads to modular forms of weight two. There are again
Jacobi sums that lead to precisely the same modular forms $f_A(q),
A=\rmI, \rmII, \rmIII$ of weight two and levels $N_A = 27, 144,
432$, possibly including a twist, as for $X_3^6$. This is expected
because $X_3^{12}$ contains the plane Fermat curve $C^6 \subset
\mathP_2$ already encountered in $X_3^6$. There is a further
L-series of a rank two motive that is determined by the quartic
elliptic weighted Fermat curve $E^4 \subset \mathP_{(1,1,2)}$
considered in Theorem 6
$$
 L_\rmIV(X_3^{12},s) ~=~ L(E^4,s-1).
 $$

\vskip .2truein

\section{String modular rigid Calabi-Yau threefolds}

The purpose of this section is to show that the modularity results
of \cite{su02, ls04} lead to a string modular interpretation of
two rigid Calabi-Yau manifolds. It has been known for a long time
in the context of lattice constructions of orbifolds that rigid
Calabi-Yau manifolds are obtained for certain quotients of
six-dimensional tori by discrete groups. In the context of zeta
functions it is more useful to consider triple products of
Brieskorn-Pham curves. Consider the elliptic Brieskorn-Pham curve
 of degree three $E^3 \subset \mathP_2$ and the quartic $E^4
 \subset \mathP_{(1,1,2)}$. On the triple products $(E^d)^3$,
 $d=3,4$ there are group actions $\mathZ_d\times \mathZ_d$,
 where $\mathZ_d = \mathZ/d\mathZ$. The quotients of the triple
 products by these groups are singular and their resolutions
 $X_3^d = \rmres((E^d)/\mathZ_d \times \mathZ_d)$ are rigid with
  \bea
   h^{1,1}(X_3^3) &=& 84 \nn \\
   h^{1,1}(X_3^4) &=& 90.
  \eea
 The string modular forms of $E^d$ are determined in terms of
  Hecke's indefinite modular forms by Theorem 6.
  Since the varieties $X_3^d$ for $d=3,4$ are rigid it is to be
 expected that they are modular and that their modular forms can
 be expressed in terms of the Hecke indefinite modular forms
 $\Theta^k_{\ell,m}(\tau)$ as well. The fact that this is indeed the case
 follows from a result shown by Cynk and Hulek, who prove in
 \cite{ch05} the following result.

 {\bf Theorem 8.}~ {\it The L-series of the rigid Calabi-Yau threefolds
  $X_3^d$, $d=3,4$ are given by cusp forms $f_d$ of weight 4
  with complex multiplication in $\mathQ(\sqrt{-d})$.}

 The modular forms $f_d$ are precisely the modular forms
 determined by the elliptic curves considered in Theorem 6. This can be
 seen as follows. Consider the character $\psi_{27}$  defined
 in Section 2 and the twisted Gr\"o\ss encharakter
 $\psi_{64} = \psi_{32}\otimes \chi_2$.
  It was shown in \cite{rs05} that the modular forms of the
  elliptic curves $E^d$ have complex multiplication in $\mathQ(\sqrt{-d})$
   and that
  \beq
   L(E^d,s) = L(\psi_{d^3},s).
  \eeq
  For $d=3$ the form $f_3 \in S_4(\G_0(9))$ is the eta product
  $\eta^8(q^3)$, which can be written in terms of Hecke's
  indefinite forms $\Theta^k_{\ell,m}$ as $f_3(q) =
  \Theta^1_{1,1}(q^3)^4$, with the associated L-series given as
  \beq
   L(X_3^3,s) = L(f_3,s) = L(\psi_{27}^3,s).
  \eeq
  For $d=4$ the form $f_4$ is given by the cube of the character
  of conductor 64
  \beq
   L(X_3^4,s) = L(\psi_{64}^3,s).
  \eeq
  It follows from these considerations that both rigid manifolds
  $X_3^d$, $d=3,4$ lead to $\Om-$motivic L-series that admit a
  string theoretic interpretation in terms of Hecke's indefinite
  modular forms.

 This result is relevant in the context of mirror symmetry. From
 the point of view of the conformal field theory on the string
 worldsheet mirror theories are isomorphic. A string theoretic
 interpretation of a geometric modular form therefore leads to the
 expectation that the motivic L-function of a variety should be
 identical to that of its mirror. This problem is considered in
 \cite{kls08} in the context of the rigid mirror manifolds
 considered above. It is shown there that one can associate mirror motives
 to the $\Om-$motives of the rigid varieties $X_3^d$, and that these
 motives are modular and lead to the same modular forms.

\vskip .2truein

\section{A modular Calabi-Yau fourfold $X_4^6 \subset \mathP_5$}

Calabi-Yau varieties of complex dimension four are useful  in the
context of F-theory in four dimensions and M-theory in three
dimensions. In this section it is shown that the $\Om-$motive of
the fourfold of degree six in projective fivespace $\mathP_5$
defined by the Brieskorn-Pham hypersurface (\ref{deg6fourfold})
 is modular.

 The Galois orbit is of length two, and the motivic L-function is
 described by computing the Jacobi sums $j_p(\a_{\Om})$ with
  $\a_{\Om} = \left(\frac{1}{6}, \dots , \frac{1}{6}\right)$.
 The results are collected in Table 7.
 \begin{center}
 \begin{small}
 \begin{tabular}{l| c c c c c c}
  $p^f$    &7  &13  &19 &25 &31  &37   \tabroom \\
 \hline

$j_{p^f}\left(\a_{\Om}\right)$
             &$-\frac{71}{2}-\frac{39}{2}\sqrt{-3}$
             &$-\frac{337}{2}-\frac{15}{2}\sqrt{-3}$
             &$\frac{601}{2}+\frac{231}{2}\sqrt{-3}$
             &625
             &$-97+552\sqrt{-3}$
             &$-\frac{529}{2} - \frac{1551}{2}\sqrt{-3}$
                  \tabroom \\

\hline
 $\b_{p^f}$      &$-71$
             &$-337$
             &601
             &625
             &$-194$
             &$-529$
               \tabroom \\

 \hline
 \end{tabular}
 \end{small}

{{\bf Table 7.}~{\it Jacobi sums for $X_4^6$.}}
 \end{center}
The resulting L-function is given by
 \beq
  L_{\Om}(X_4^6,s) \doteq  1 - \frac{71}{7^s} - \frac{337}{13^s} +
      \frac{601}{19^s} + \frac{625}{25^s} - \frac{194}{31^s} - \frac{529}{37^s} + \cdots
 \eeq
 The associated $q-$expansion $f_{\Om}(X_4^6,q)$ differs from the
 that of a newform of weight five and level 27
 \beq
  f_{27}(q) = q + 71q^7 - 337q^{13} - 601q^{19}
  + 194q^{31} - 529q^{37} + \cdots
 \eeq
 only in signs. These signs can be adjusted the quadratic character
 $\chi_3$ defined by the Legendre symbol
  \beq
   \chi_3(p) = \left(\frac{3}{p}\right),
  \eeq
  leading to
  \beq
   f_{\Om}(X_4^6,q) = f_{27}(q) \otimes \chi_3.
  \eeq

 This modular form can also be described as a
 Hecke L-series associated to the character
  $\psi_{27}$ associated to the Eisenstein field
  $K=\mathQ(\sqrt{-3})$ and defined in
  Section 2 in (\ref{eisenstein-hecke-chars}).
 More precisely, the twisted Hecke L-series agrees with that
 of the $\Om-$motive of $X_4^6$
  \beq
   L_{\Om}(X_4^6,s) = L(\psi_{27}^4,s) \otimes \chi_3.
  \eeq
 Hence the motivic L-function is again a purely algebraic object
 and its fundamental structure is determined by the L-series of a
  Hecke indefinite theta series as noted in (\ref{string-e3}).

\vskip .2truein

\section{Emergent space from characters and modular forms}

In the discussion so far the goal was to formulate a general
framework of $\Om-$motives of varieties of Calabi-Yau type, and
more generally, of special Fano type, in the context
Grothendieck's framework of motives, and to test the conjecture
that these motives are string modular in the sense that it is
possible to identify modular forms on the worldsheet whose Mellin
transform agrees with the L-function of the resulting motives. The
problem of constructing spacetime geometry from fundamental string
input involves the inverse problem of this strategy.

The aim of the present section is to address this "space
construction" problem. As already mentioned earlier, the idea here
must be to obtain a construction of the motivic pieces of the
compact varieties from the basic conformal field theoretic modular
forms. There are several ways to think about these objects and the
following remarks describe how these various constructions, which
a priori are independent, fit together in the context of the
$\Om-$motive of Brieskorn-Pham type varieties.

The main simplifying observation in the present context of
weighted Fermat hypersurfaces is that all the motivic L-functions
$L_{\Om}(X,s)$ that have been encountered so far in the program
initiated in \cite{rs01, su02} and continued in \cite{ls04, rs05,
rs06, rs07} lead to modular forms which are of complex
multiplication type \cite{r77} (see \cite{y06} for geometric
aspects of CM). The structure of such forms has been described in
detail in \cite{lss03,rs05,rs06}. The important point in the
present context is that modular $\Om-$motives of CM type are
algebraic objects whose L-functions are given by the Hecke
L-series of a Gr\"o\ss encharakter (possibly modulo a twist). This
construction of algebraic Hecke characters from geometry can be
inverted, and it is known how to construct motives directly from
the characters. It is in particular possible to construct a
Grothendieck motive of the form $M_{\chi} = (A_{\chi},p_{\chi})$,
where $A_{\chi}$ is an abelian variety associated to the character
$\chi$ by the theorem of Casselman \cite{gs71a}, and $p_{\chi}$ is
a projector associated to $\chi$. This leads to an apparent
problem of riches, because given any modular form $S_w(\G_0(N))$
it is possible to construct a Grothendieck motive $M_f = (X_f,
p_f)$ by considering the cohomology of an associated Kuga-Sato
variety $X_f$, as shown by Deligne \cite{d69}, Jannsen \cite{j90},
and Scholl \cite{as90}. Combining the abelian motives and the
Kuga-Sato motives with those of the $\Om-$motives thus leads to
three a priori different motivic constructions associated to the
algebraic Hecke characters encountered here and in the earlier
papers.

It turns out that the motives $M_{\chi}$, $M_f$ and $M_{\Om}$ all
are isomorphic because they arise from the same CM modular form.
This follows because motives associated to CM modular forms have
CM, and one can generalize Faltings result that L-series
characterize abelian varieties up to isogeny to CM motives, as
shown e.g. by Anderson \cite{a86}.

In the case of varieties of Brieskorn-Pham type it is possible to
make the relation between abelian varieties and $\Om-$motives more
concrete by noting that the cohomology of Fermat varieties has a
well-known inductive structure which was first noted by
Shioda-Katsura \cite{sk79} (see also Deligne \cite{d82}) in the
context of Fermat varieties. This inductive structure allows to
reduce the cohomology of higherdimensional varieties in terms of
the cohomology of algebraic curves (modulo Tate twists). Hence the
basic building blocks are abelian varieties derived from the
Jacobians of these curves. It follows from results of Gross and
Rohrlich \cite{gr78} that these Jacobians factor into simple
abelian varieties and that these abelian factors have complex
multiplication. The final step in the construction is provided by
the fact that the L-function of the abelian variety attached to
$\chi$ by Casselman's result is given by the conjugates of the
L-function of the Hecke character.

\vskip .2truein

\section{Further considerations}

The goal of the program continued in this paper is to investigate
the relation between the geometry of spacetime and the physics of
the worldsheet by analyzing in some depth the connection between
the modular symmetry encoded in exact models on the worldsheet and
the modular symmetries that emerge from the nontrivial arithmetic
structure of spacetime.  The techniques introduced for this
purpose provide a stronger, and more precise, alternative to the
framework of Landau-Ginzburg theories and $\si-$models. The latter
in particular presupposes the concept of an ambient space in which
the string propagates, a notion that should emerge as a derived
concept in a fundamental theory.

The focus of the results obtained in previous work and the present
paper has been on the class of diagonal models, given by Gepner's
construction. It would be interesting to extend these
considerations to the more general class of Kazama-Suzuki models
\cite{ks89}. Of particular interest in that class are certain
'irreducible' models which are not tensor products, hence a single
conformal field theoretic quotient suffices to saturate the
necessary central charge. Such models exist for both K3 surfaces
and Calabi-Yau threefolds, and establishing modularity in the
sense described here would be a starting point for the exploration
of modular points in the moduli space of nondiagonal varieties.
Results in this direction would illuminate relations between
different conformal field theories.

A second open problem is the analysis of families of varieties
with respect to their modular properties. First steps in this
direction have been taken in refs. \cite{cdr00, cdr03, k04} where
the zeta functions for particular one-parameter families of
Calabi-Yau threefolds are computed. It would be of interest to
understand how the modular behavior of these families is related
to deformations along marginal directions of the associated
conformal field theory. Such an analysis might be useful for the
understanding of the conformal field theoretic behavior of
conifold phase transitions.

\vskip .3truein

 {\bf ACKNOWLEDGEMENT.}

It is a pleasure to thank Monika Lynker for conversations, and Rob
Myers for raising the point of string dualities in the context of
the motivic picture. Part of this work was completed while the
author was supported as a Scholar at the Kavli Institute for
Theoretical Physics in Santa Barbara. This work was supported in
part by the National Science Foundation under Grant No.
PHY05-51164, a KSU Incentive Grant for Scholarship, and IUSB
Faculty Research Grants.

\vskip .3truein

\baselineskip=17pt

\begin{small}

\end{small}

\end{document}